\documentclass[aps,10pt,preprint,floatfix]{revtex4-1} %
\usepackage{graphicx,color,pict2e}%
\definecolor{aliceblue}{rgb}{0.94,0.97,1.00}
\definecolor{antiquewhite}{rgb}{0.98,0.92,0.84}
\definecolor{aqua}{rgb}{0.00,1.00,1.00}
\definecolor{aquamarine}{rgb}{0.50,1.00,0.83}
\definecolor{azure}{rgb}{0.94,1.00,1.00}
\definecolor{beige}{rgb}{0.96,0.96,0.86}
\definecolor{bisque}{rgb}{1.00,0.89,0.77}
\definecolor{black}{rgb}{0.00,0.00,0.00}
\definecolor{blanchedalmond}{rgb}{1.00,0.92,0.80}
\definecolor{blue}{rgb}{0.00,0.00,1.00}
\definecolor{blueviolet}{rgb}{0.54,0.17,0.89}
\definecolor{brass}{rgb}{0.71,0.65,0.26}
\definecolor{brown}{rgb}{0.65,0.16,0.16}
\definecolor{burlywood}{rgb}{0.87,0.72,0.53}
\definecolor{cadetblue}{rgb}{0.37,0.62,0.63}
\definecolor{chartreuse}{rgb}{0.50,1.00,0.00}
\definecolor{chocolate}{rgb}{0.82,0.41,0.12}
\definecolor{coolcopper}{rgb}{0.85,0.53,0.10}
\definecolor{copper}{rgb}{0.75,0.00,0.87}
\definecolor{coral}{rgb}{1.00,0.50,0.31}
\definecolor{cornflower}{rgb}{0.75,0.94,0.87}
\definecolor{cornflowerblue}{rgb}{0.39,0.58,0.93}
\definecolor{cornsilk}{rgb}{1.00,0.97,0.86}
\definecolor{crimson}{rgb}{0.86,0.08,0.24}
\definecolor{cyan}{rgb}{0.00,1.00,1.00}
\definecolor{darkblue}{rgb}{0.00,0.00,0.55}
\definecolor{darkbrown}{rgb}{0.85,0.04,0.00}
\definecolor{darkcyan}{rgb}{0.00,0.55,0.55}
\definecolor{darkgoldenrod}{rgb}{0.72,0.53,0.04}
\definecolor{darkgray}{rgb}{0.66,0.66,0.66}
\definecolor{darkgreen}{rgb}{0.00,0.39,0.00}
\definecolor{darkkhaki}{rgb}{0.74,0.72,0.42}
\definecolor{darkmagenta}{rgb}{0.55,0.00,0.55}
\definecolor{darkolivegreen}{rgb}{0.33,0.42,0.18}
\definecolor{darkorange}{rgb}{1.00,0.55,0.00}
\definecolor{darkorchid}{rgb}{0.60,0.20,0.80}
\definecolor{darkred}{rgb}{0.55,0.00,0.00}
\definecolor{darksalmon}{rgb}{0.91,0.59,0.48}
\definecolor{darkseagreen}{rgb}{0.56,0.74,0.56}
\definecolor{darkslateblue}{rgb}{0.28,0.24,0.55}
\definecolor{darkslategray}{rgb}{0.18,0.31,0.31}
\definecolor{darkturquoise}{rgb}{0.00,0.81,0.82}
\definecolor{darkviolet}{rgb}{0.58,0.00,0.83}
\definecolor{deeppink}{rgb}{1.00,0.08,0.58}
\definecolor{deepskyblue}{rgb}{0.00,0.75,1.00}
\definecolor{dimgray}{rgb}{0.41,0.41,0.41}
\definecolor{dodgerblue}{rgb}{0.12,0.56,1.00}
\definecolor{feldsper}{rgb}{1.00,0.82,0.88}
\definecolor{firebrick}{rgb}{0.70,0.13,0.13}
\definecolor{floralwhite}{rgb}{1.00,0.98,0.94}
\definecolor{forestgreen}{rgb}{0.13,0.55,0.13}
\definecolor{fuchsia}{rgb}{1.00,0.00,1.00}
\definecolor{gainsboro}{rgb}{0.86,0.86,0.86}
\definecolor{ghostwhite}{rgb}{0.97,0.97,1.00}
\definecolor{gold}{rgb}{1.00,0.84,0.00}
\definecolor{goldenrod}{rgb}{0.85,0.65,0.13}
\definecolor{gray}{rgb}{0.50,0.50,0.50}
\definecolor{green}{rgb}{0.00,0.50,0.00}
\definecolor{greenyellow}{rgb}{0.68,1.00,0.18}
\definecolor{honeydew}{rgb}{0.94,1.00,0.94}
\definecolor{hotpink}{rgb}{1.00,0.41,0.71}
\definecolor{indianred}{rgb}{0.80,0.36,0.36}
\definecolor{indigo}{rgb}{0.29,0.00,0.51}
\definecolor{ivory}{rgb}{1.00,1.00,0.94}
\definecolor{khaki}{rgb}{0.94,0.90,0.55}
\definecolor{lavender}{rgb}{0.90,0.90,0.98}
\definecolor{lavenderblush}{rgb}{1.00,0.94,0.96}
\definecolor{lawngreen}{rgb}{0.49,0.99,0.00}
\definecolor{lemonchiffon}{rgb}{1.00,0.98,0.80}
\definecolor{lightblue}{rgb}{0.68,0.85,0.90}
\definecolor{lightcoral}{rgb}{0.94,0.50,0.50}
\definecolor{lightcyan}{rgb}{0.88,1.00,1.00}
\definecolor{lightgoldenrodyellow}{rgb}{0.98,0.98,0.82}
\definecolor{lightgreen}{rgb}{0.56,0.93,0.56}
\definecolor{lightgrey}{rgb}{0.83,0.83,0.83}
\definecolor{lightpink}{rgb}{1.00,0.71,0.76}
\definecolor{lightsalmon}{rgb}{1.00,0.63,0.48}
\definecolor{lightseagreen}{rgb}{0.13,0.70,0.67}
\definecolor{lightskyblue}{rgb}{0.53,0.81,0.98}
\definecolor{lightslategray}{rgb}{0.47,0.53,0.60}
\definecolor{lightsteelblue}{rgb}{0.69,0.77,0.87}
\definecolor{lightyellow}{rgb}{1.00,1.00,0.88}
\definecolor{lime}{rgb}{0.00,1.00,0.00}
\definecolor{limegreen}{rgb}{0.20,0.80,0.20}
\definecolor{linen}{rgb}{0.98,0.94,0.90}
\definecolor{magenta}{rgb}{1.00,0.00,1.00}
\definecolor{maroon}{rgb}{0.50,0.00,0.00}
\definecolor{mediumaquamarine}{rgb}{0.40,0.80,0.67}
\definecolor{mediumblue}{rgb}{0.00,0.00,0.80}
\definecolor{mediumorchid}{rgb}{0.73,0.33,0.83}
\definecolor{mediumpurple}{rgb}{0.58,0.44,0.86}
\definecolor{mediumseagreen}{rgb}{0.24,0.70,0.44}
\definecolor{mediumslateblue}{rgb}{0.48,0.41,0.93}
\definecolor{mediumspringgreen}{rgb}{0.00,0.98,0.60}
\definecolor{mediumturquoise}{rgb}{0.28,0.82,0.80}
\definecolor{mediumvioletred}{rgb}{0.78,0.08,0.52}
\definecolor{midnightblue}{rgb}{0.10,0.10,0.44}
\definecolor{mintcream}{rgb}{0.96,1.00,0.98}
\definecolor{mistyrose}{rgb}{1.00,0.89,0.88}
\definecolor{moccasin}{rgb}{1.00,0.89,0.71}
\definecolor{navajowhite}{rgb}{1.00,0.87,0.68}
\definecolor{navy}{rgb}{0.00,0.00,0.50}
\definecolor{oldlace}{rgb}{0.99,0.96,0.90}
\definecolor{olive}{rgb}{0.50,0.50,0.00}
\definecolor{olivedrab}{rgb}{0.42,0.56,0.14}
\definecolor{orange}{rgb}{1.00,0.65,0.00}
\definecolor{orangered}{rgb}{1.00,0.27,0.00}
\definecolor{orchid}{rgb}{0.85,0.44,0.84}
\definecolor{palegoldenrod}{rgb}{0.93,0.91,0.67}
\definecolor{palegreen}{rgb}{0.60,0.98,0.60}
\definecolor{paleturquoise}{rgb}{0.69,0.93,0.93}
\definecolor{palevioletred}{rgb}{0.86,0.44,0.58}
\definecolor{papayawhip}{rgb}{1.00,0.94,0.84}
\definecolor{peachpuff}{rgb}{1.00,0.85,0.73}
\definecolor{peru}{rgb}{0.80,0.52,0.25}
\definecolor{pink}{rgb}{1.00,0.75,0.80}
\definecolor{plum}{rgb}{0.87,0.63,0.87}
\definecolor{powderblue}{rgb}{0.69,0.88,0.90}
\definecolor{purple}{rgb}{0.50,0.00,0.50}
\definecolor{red}{rgb}{1.00,0.00,0.00}
\definecolor{richblue}{rgb}{0.05,0.69,0.88}
\definecolor{rosybrown}{rgb}{0.74,0.56,0.56}
\definecolor{royalblue}{rgb}{0.25,0.41,0.88}
\definecolor{saddlebrown}{rgb}{0.55,0.27,0.07}
\definecolor{salmon}{rgb}{0.98,0.50,0.45}
\definecolor{sandybrown}{rgb}{0.96,0.64,0.38}
\definecolor{seagreen}{rgb}{0.18,0.55,0.34}
\definecolor{seashell}{rgb}{1.00,0.96,0.93}
\definecolor{sienna}{rgb}{0.63,0.32,0.18}
\definecolor{silver}{rgb}{0.75,0.75,0.75}
\definecolor{skyblue}{rgb}{0.53,0.81,0.92}
\definecolor{slateblue}{rgb}{0.42,0.35,0.80}
\definecolor{slategray}{rgb}{0.44,0.50,0.56}
\definecolor{snow}{rgb}{1.00,0.98,0.98}
\definecolor{springgreen}{rgb}{0.00,1.00,0.50}
\definecolor{steelblue}{rgb}{0.27,0.51,0.71}
\definecolor{ctan}{rgb}{0.82,0.71,0.55}
\definecolor{teal}{rgb}{0.00,0.50,0.50}
\definecolor{thistle}{rgb}{0.85,0.75,0.85}
\definecolor{tomato}{rgb}{1.00,0.39,0.28}
\definecolor{turquoise}{rgb}{0.25,0.88,0.82}
\definecolor{violet}{rgb}{0.93,0.51,0.93}
\definecolor{wheat}{rgb}{0.96,0.87,0.70}
\definecolor{white}{rgb}{1.00,1.00,1.00}
\definecolor{whitesmoke}{rgb}{0.96,0.96,0.96}
\definecolor{yellow}{rgb}{1.00,1.00,0.00}
\definecolor{yellowgreen}{rgb}{0.60,0.80,0.20}

\usepackage{dcolumn}%
\usepackage{bm}%
\usepackage{amsmath,amsthm,amssymb}
\usepackage{braket,here}
\usepackage{bigstrut}
\usepackage{multirow}
\usepackage{comment}
\usepackage{autobreak}

\vbadness=\maxdimen
\allowdisplaybreaks[4]
\newcount\hbadness
\newdimen\hfuzz

\def\D{\mathrm{d}}
\def\E{\mathrm{e}}

\def\abs#1{{\left\rvert #1 \right\lvert}}

\DeclareMathOperator{\Tr}{Tr}

\begin{document}
\title{%
Non-collinearity Effects on Magnetocrystalline Anisotropy for $R_2$Fe$_{14}$B Magnets
}

\author{Daisuke Miura}
\affiliation{%
Department of Applied Physics, Tohoku University, Sendai 980-8579, Japan}
\email{dmiura@solid.apph.tohoku.ac.jp}

\author{Akimasa Sakuma}
\affiliation{Department of Applied Physics, Tohoku University, Sendai 980-8579, Japan}

\date{\today}

\begin{abstract}
We present a theoretical investigation of the magnetocrystalline anisotropy (MA)
in $R_2$Fe$_{14}$B ($R$ is a rare-earth element) magnets in consideration of the non-collinearity effect (NCE)
between the $R$ and Fe magnetization directions.
In particular, the temperature dependence of the MA of Dy$_2$Fe$_{14}$B magnets is detailed in terms of the $n$th-order MA constant (MAC)
$K_n(T)$
at a temperature $T$.
The features of this constant are as follows:
$K_1(T)$ has a broad plateau in the low-temperature range
and
$K_2(T)$ persistently survives in the high-temperature range.
The present theory explains these features in terms of the NCE on the MA
by using numerical calculations for the entire temperature range,
and further, by using a high-temperature expansion.
The high-temperature expansion for $K_n(T)$ is expressed in the form of
$K_n(T)=\kappa_1(T)\left[1+\delta(T)\right][-\delta(T)]^{n-1}$,
where $\kappa_1(T)$ is the part without the NCE
and $\delta(T)$ is a correction factor for the NCE introduced in this study.
We also provide a convenient expression to evaluate $K_n(T)$,
which can be determined only by a second-order crystalline electric field coefficient and an effective exchange field.
\end{abstract}

\maketitle

\section{Introduction}
\label{intro}

$R_2$Fe$_{14}$B compounds ($R$ is mainly a rare-earth element) have been research targets in the fields of not only engineering but also science;
specifically, the magnetism of these compounds has been systematically investigated both experimentally and theoretically\cite{Herbst1991,Skomski1999,Kuzmin2007,Coey2010,Hirosawa2017,Miyake2018}.
Present-day high-performance computers allow us
to directly calculate these electronic and/or magnetic structures in complex and large calculation models for $R_2$Fe$_{14}$B-based systems,
which also becomes a motivation for developing new numerical methods such as
high-accuracy first-principles calculation methods\cite{Miyake2018},
constrained Monte Carlo methods\cite{Asselin2010},
and
finite-temperature Landau--Lifshitz--Gilbert analyses\cite{Nishino2015,Nishino2018}.
Furthermore, the fields of engineering
strongly require
theoretical guidelines to develop magnets, the performance of which exceeds that of Nd-Fe-B magnets.
Most recently,
the above-mentioned numerical methods have been applied to a realistic model for rare-earth intermetallics,
and quantitative results comparable to the experimental results were obtained
by first-principles calculations\cite{Miura2014d,Yoshioka2015,Tatetsu2016,Tatetsu2018a,Yoshioka2018,Tsuchiura2018a}
and by Monte Carlo methods\cite{Toga2016,Matsumoto2016,Nishino2017,Toga2018,Westmoreland2018}.
On the other hand, to date,
simpler analyses have also been conducted
on the basis of
phenomenological theory\cite{Skomski1998,Skomski2006,Skomski2009,Miura2018}
or
mean field theory (MFT)\cite{Kuzmin2007,Sasaki2015,Miura2015,Ito2016,Yoshioka2018}
to understand the mechanism of the coercive forces of rare-earth permanent magnets and to identify the factors dominating these mechanisms.

The magnetocrystalline anisotropy (MA) of a magnet refers to its free energy density as a function of the magnetization direction.
In \textit{simple} magnets\cite{Skomski2013},
the free energy density is well expressed by the single term $K_1(T)\sin^2\Theta$, where $K_1(T)$ is the first-order MA constant (MAC)
and $\Theta$ is the zenithal angle of the magnetization measured from the crystal axis.
However, in rare-earth (RE) magnets, the angle dependence of the free energy density has a more complex form,
especially in the low-temperature range\cite{Hirosawa1986,Durst1986,Sagawa1987,Cadogan1988};
assuming tetragonal symmetry such as that of $R_2$Fe$_{14}$B compounds,
the free energy density can be expressed as 
\begin{align}
\sum_{m=0}^\infty
\sum_{n=0}^{\lfloor m/2 \rfloor}
K_m^{(\prime)^n}(T)\sin^{2m}\Theta \cos 4n\Phi
,\label{eq:MACs}
\end{align}
where $K_m^{(\prime)^n}(T)$ is the $m$th-order MAC
and $\Phi$ is the azimuthal angle of the magnetization.
The expansion presents
a convergence problem and it is difficult to uniquely determine $K_m^{(\prime)^n}(T)$ because of the non-orthogonality of the basis set of $\set{\sin^{2m}\Theta \cos 4n\Phi}$ as reviewed by Kuz'min\cite{Kuzmin1995}.
Thus, MACs have been evaluated by assuming convergence or by using fitting methods that assume a finite-expansion form for practical purposes
\cite{Hirosawa1985a,
Hirosawa1985,
Yamada1986,
Hirosawa1986,
Grossinger1986,
Durst1986,
Hirosawa1987,
Sagawa1987,
Otani1987,
Cadogan1988,
Radwanski1989}.

In our previous studies on the MA of Nd$_2$Fe$_{14}$B magnets\cite{Sasaki2015,Miura2015,Miura2018},
the total magnetization was assumed to be collinear to the Fe magnetization.
However, this assumption raises a serious error in evaluations for the MA of the $R_2$Fe$_{14}$B magnet,
the $R$ magnetization of which is highly non-collinear to its Fe magnetization.
For example, Dy$_2$Fe$_{14}$B magnets exhibit the non-collinearity effect (NCE) on the temperature dependence of the MA.
Recently,
Ito et al.\cite{Ito2016} have calculated $K_1(T)$ of Dy$_2$Fe$_{14}$B magnets
without the NCE,
and then they demonstrated that the resultant $K_1(T)$ rapidly decays with increasing temperature;
however, as is well known in experiments, $K_1(T)$ of Dy$_2$Fe$_{14}$B magnets has a broad plateau in the low-temperature range.
In addition, they pointed out the importance of the NCE via an MFT analysis of the magnetization curves of Dy$_2$Fe$_{14}$B magnets.
In Sect. \ref{NCE}, we clearly show that the NCE on $K_1(T)$ is the origin of the disagreement between the MFT and the experiments on Dy$_2$Fe$_{14}$B magnets.

In this study,
we theoretically investigated the NCE on the temperature-dependent MA of $R_2$Fe$_{14}$B magnets
by using numerical calculations for the entire temperature range.
Furthermore,
in the high-temperature range,
we provide explicit expressions to describe the NCE
and clarify our understanding of how the NCE appears in the MA.
We also provide a practical expression to estimate the temperature dependence of the MA in RE intermetallics.
The present article is constructed as follows:
In Sect. \ref{PU}, we review previous theoretical work on the temperature dependence of $K_n(T)$ without NCEs in RE intermetallics.
In Sect. \ref{NCE}, we show how the NCEs on the MA appear by taking the Nd$_2$Fe$_{14}$B and Dy$_2$Fe$_{14}$B magnets as examples,
and we develop microscopic expressions for the MA with NCEs in $R_2$Fe$_{14}$B magnets in the high-temperature range.
In addition, we apply the results to $R$=Tb, Dy, Ho, Er, Tm, and Yb.
In Sect. \ref{summary}, we summarize this study.

\section{Present Understanding of the Temperature-dependent MA in $R_2\mathrm{Fe}_{14}\mathrm{B}$ Magnets without NCEs}
\label{PU}

First, let us briefly recall important theoretical studies on temperature-dependent MA in two-sublattice systems.
Although it is difficult to directly express the temperature-dependent MACs under general conditions\cite{Kuzmin1995},
several explicit expressions have been obtained for limited situations.
Here, we consider the temperature-dependent MA of an $R_2$Fe$_{14}$B magnet as an example of the two-sublattice system.
Assuming that the Fe magnetization is collinear to the total magnetization in the magnet,
we can evaluate the MACs from the MA as a function of the Fe magnetization angle,
and then the total $m$th-order MAC is separated into the $R$- and Fe-sublattice contributions as
\begin{align}
K_m^{(\prime)^n}(T)=
K_m^{R(\prime)^n}(T)
+
K_m^{\mathrm{Fe}(\prime)^n}(T)
.
\end{align}
Here, we assume that $K_m^{\mathrm{Fe}(\prime)^n}(T)$ is obtained from the experimental results
for $R_2$Fe$_{14}$B magnets with a nonmagnetic $R$ element such as $R=$Y,
and therefore, we focus only on $K_m^{R(\prime)^n}(T)$.
A qualitative (but simple) understanding of $K_m^{R(\prime)^n}(T)$ can be obtained
from the power-law scenario derived by Zener\cite{Zener1954};
most recently\cite{Miura2018},
we have explicitly expressed the extended form of the Akulov--Zener--Callen--Callen law\cite{Akulov1936,Zener1954,Callen1966}
(or today simply known as the Callen--Callen law)
up to the third order, as
\begin{subequations}
\begin{align}
&K_1^R(T)=K_1^R(0)\mu_R(T)^3
\notag\\&
+\frac{8}{7}K_2^R(0)\left[\mu_R(T)^3-\mu_R(T)^{10}\right]\notag\\&+\frac{8}{7}K_3^R(0)\left(\mu_R(T)^3-\frac{18}{11}\mu_R(T)^{10}+\frac{7}{11}\mu_R(T)^{21}\right),
\label{eq:K1}
\\
&K_2^R(T)=K_2^R(0)\mu_R(T)^{10}
\notag\\&
+\frac{18}{11}K_3^R(0)\left[\mu_R(T)^{10}-\mu_R(T)^{21}\right],
\label{eq:K2}
\\
&K_2^{R\prime}(T)=K_2^{R\prime}(0)\mu_R(T)^{10}
\notag\\&
+\frac{10}{11}K_3^{R\prime}(0)\left[\mu_R(T)^{10}-\mu_R(T)^{21}\right],
\\
&K_3^R(T)=K_3^R(0)\mu_R(T)^{21},
\label{eq:K3}
\\
&K_3^{R\prime}(T)=K_3^{R\prime}(0)\mu_R(T)^{21},
\end{align}
\label{eq:extend}%
\end{subequations}
where $\mu_R(T)$ is the normalized $R$ magnetization with $\mu_R(0)=1$,
and we demonstrated that the experimental results for Nd$_2$Fe$_{14}$B magnets well obey the extended Callen--Callen law [Eq. (\ref{eq:extend})].
This view of the power law enables us
to immediately establish
that a narrow plateau appears in the low-temperature range
and
that higher-order MACs rapidly decay with increasing temperature compared with lower-order ones.
These features describe the general behavior of on-site MA in homogeneous local moment systems\cite{Kuzmin1992,Skomski1998,Ito2016,Miura2018}.

On the other hand,
to reflect the material individuality,
a microscopic description for the MA is appropriate.
Many authors have reported the microscopic theory for temperature-dependent MACs\cite{Herbst1991,Kuzmin2007}.
At zero temperature,
Yamada et al.\cite{Yamada1988} reported the explicit relation between MACs and crystalline electric fields (CEFs)
under the conditions of a weak CEF, strong effective exchange field (EXF), and strong spin-orbit interaction (SOI) (i.e., only the ground $J$ multiplet is considered) on the $R$ sites:
\begin{subequations}
\begin{align}
K_1^R(0)&= -3 f_2 B_2^0-40 f_4 B_4^0-168 f_6 B_6^0,\label{eq:YKYN0}\\
K_2^R(0)&= 35 f_4 B_4^0+378f_6 B_6^0,\\
K_2^{R\prime}(0)&= f_4 B_4^4+10 f_6 B_6^4,\\
K_3^R(0)&= -231 f_6 B_6^0,\\
K_3^{R\prime}(0)&= -11 f_6 B_6^4,
\end{align}
\label{eq:YKYN}%
\end{subequations}
where $f_k:=2^{-k}(2J)!/(2J-k)!$ and $B_k^q$ is the CEF coefficient.
Under the same conditions except at zero temperature,
in 1992,
Kuz'min reported directly comparable results with the power law [Eq. (\ref{eq:extend})]
as
\def\BJ#1{%
\tilde{\mathrm{B}}_J^#1(x)
}
\begin{subequations}
\begin{align}
&K_1^R(T)=K_1^R(0)
\BJ{2}
\notag\\&
+\frac{8}{7}K_2^R(0)\left[
\BJ{2}
-
\BJ{4}
\right]
\notag\\
&+\frac{8}{7}K_3^R(0)\left(\BJ{2}-\frac{18}{11}\BJ{4}+\frac{7}{11}\BJ{6}\right),
\label{eq:K1CEF}
\\
&K_2^R(T)=K_2^R(0)\BJ{4}
\notag\\&
+\frac{18}{11}K_3^R(0)\left[\BJ{4}-\BJ{6}\right],
\label{eq:K2CEF}
\\
&K_2^{R\prime}(T)=K_2^{R\prime}(0)\BJ{4}
\notag\\&
+\frac{10}{11}K_3^{R\prime}(0)\left[\BJ{4}-\BJ{6}\right],
\\
&K_3^R(T)=K_3^R(0)\BJ{6},
\label{eq:K3CEF}
\\
&K_3^{R\prime}(T)=K_3^{R\prime}(0)\BJ{6},
\end{align}
\label{eq:extendCEF}%
\end{subequations}
where
$
\BJ{k}:=J^k\mathrm{B}_J^k(x)/f_k
$
and $\mathrm{B}_J^k(x)$ is Kuz'min's generalized Brillouin function (GBF)\cite{Kuzmin1992,Kuz1995,Kuzmin1996,Kuzmin2007}
with $x:=2J\abs{g-1} H_\mathrm{exf}(T)/(k_\mathrm{B}T)$,
where $g$ is the Land\`e $g$ factor,
$H_\mathrm{exf}(T)$ is the magnitude of the EXF,
and $k_\mathrm{B}$ is Boltzmann's constant.
Kuz'min derived the relation between the GBF and the Akulov--Zener power law for the low-temperature range:
\begin{align}
\BJ{k}\simeq \mu_R(T)^{k(k+1)/2}.
\label{eq:power}
\end{align}
This approximation was also referred to in terms of MFT by Keffer in 1955\cite{Keffer1955,Callen1966}.
Although the Akulov--Zener power law is no longer quantitatively supported by microscopic theory in the high-temperature range,
it has been confirmed that Eq. (\ref{eq:power}) is qualitatively satisfied \cite{Kuzmin1992,Ito2016}.
The reason for this finding is simple:
both monotonically decrease
and
take the same values at $T=0$ (both are 1) and $T=T_\mathrm{C}$ (both are 0),
where $T_\mathrm{C}$ is the Curie temperature.
Here, it is necessary to note that Kazakov and Andreeva\cite{Kazakov1970} derived results equivalent to Eq. (\ref{eq:extendCEF}) in 1970
(see Ref. 11 in Ref. \citenum{Kuzmin1995}).

If the low-angle limit $\Theta\to 0$ is considered, then the series in Eq. (\ref{eq:MACs}) converges,
and therefore,
it becomes possible to obtain the exact expressions for temperature-dependent MACs.
We have recently derived these expressions and applied them to the case of Nd$_2$Fe$_{14}$B\cite{Miura2015},
where the expressions reproduce Eq. (\ref{eq:extendCEF}) within the limits of the strong EXF and SOI.

As seen above,
the expressions for the temperature-dependent MA are connected under appropriate conditions,
although some expressions have been reported in different forms.
One of our aims is to reflect the NCEs in the previous results,
and this work is presented in Sect. \ref{PE}.

\section{Non-collinearity Effects on MA of $R_2\mathrm{Fe}_{14}\mathrm{B}$ Magnets}
\label{NCE}

In this section, we reveal the NCEs on the MA of $R_2\mathrm{Fe}_{14}\mathrm{B}$ magnets
on the basis of a standard ligand-field theory.
First, we define the theoretical model used in this study.

The crystal structure of $R_2$Fe$_{14}$B compounds is tetragonal with $P4_2/mnm$,
and there are eight $R$ ions in the unit cell.
The $R$ ion sites are classified into two types: $i=$ f or g. These two types are distinguished crystallographically,
and therefore the CEF Hamiltonian for the 4f electrons is written as $\mathcal{V}_\mathrm{CEF}^i$ depending on $i$\cite{Herbst1991,Haskel2005,Yoshioka2018,Tsuchiura2018a}.
Here, we consider the 4f electrons described by
\begin{align}
\mathcal{H}^i(\theta,\phi;T)
&:=\mathcal{V}_\mathrm{CEF}^i+\lambda\vec{\mathcal{L}}\cdot\vec{\mathcal{S}}-2\vec{\mathcal{S}}\cdot\bm H_\mathrm{EXF}(\theta,\phi;T),
\label{eq:hamil}
\end{align}
where
$\lambda$ is the strength of the SOI
and $\vec{\mathcal{L}} (\vec{\mathcal{S}})$ is the operator of the total angular (spin) momentum of the 4f electrons.
The EXF is assumed to be proportional to the Fe magnetization given by
\begin{align}
\bm M_\mathrm{Fe}(\theta,\phi;T)&:=M_\mathrm{Fe}\mu_\mathrm{Fe}(T)\bm m_\mathrm{Fe}(\theta,\phi),\\
\bm m_\mathrm{Fe}(\theta,\phi)&:=(\sin\theta\cos\phi,\sin\theta\sin\phi,\cos\theta);
\end{align}
that is,
\begin{align}
\bm H_\mathrm{EXF}(\theta,\phi;T)&:=H_\mathrm{EXF}\mu_\mathrm{Fe}(T)\bm m_\mathrm{Fe}(\theta,\phi),
\label{eq:EXF2}
\end{align}
where $M_\mathrm{Fe}$ and $H_\mathrm{EXF}$, respectively, are the saturation magnetization and the strength of the EXF at zero temperature,
and $\mu_\mathrm{Fe}(T)$ describes the temperature dependence of the magnitude of the Fe magnetization.

\begin{figure}[tb]
\centering
\includegraphics[width=0.3\textwidth]{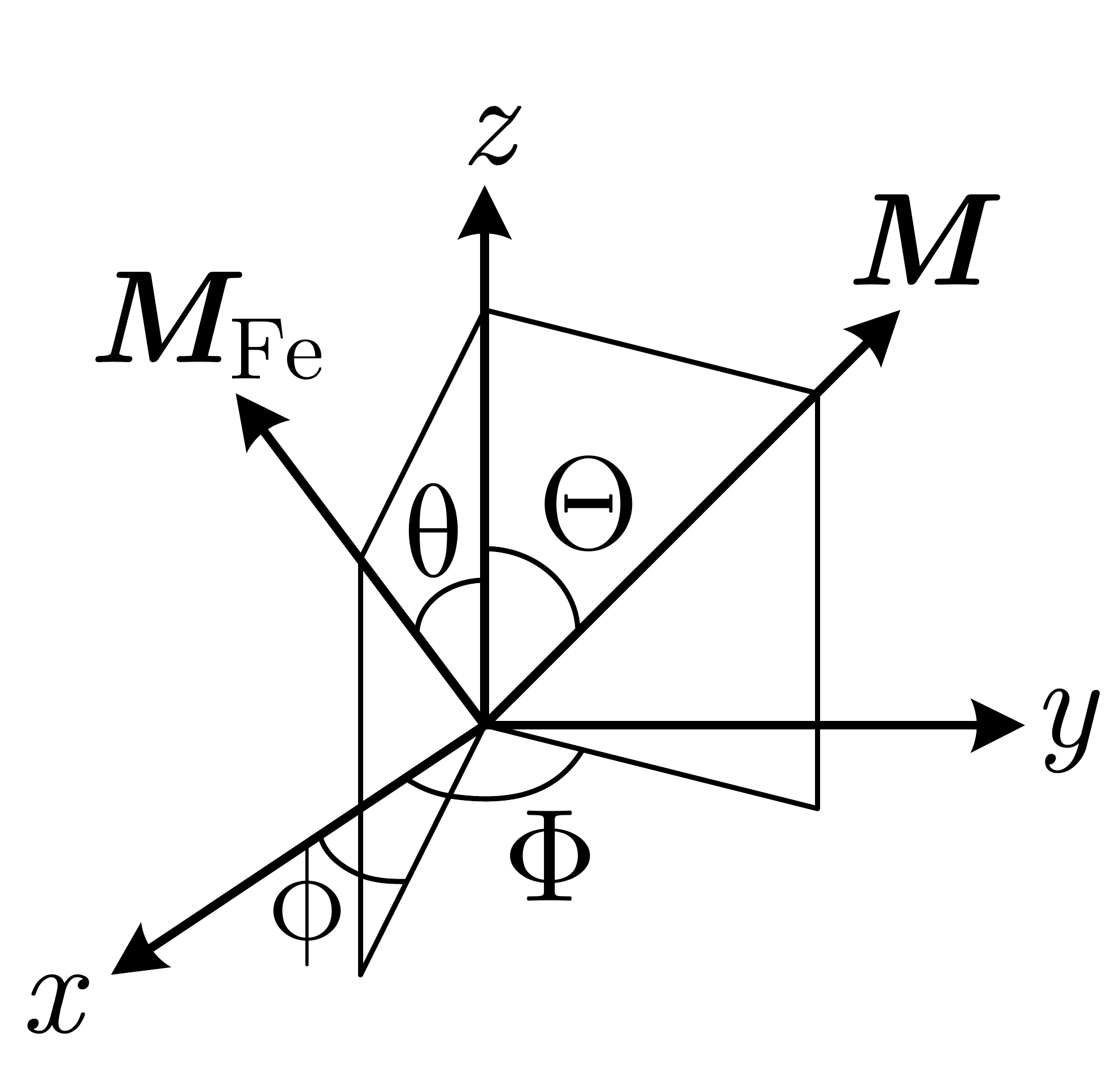}
\caption{%
Definitions of the zenithal and azimuthal angles
of the Fe magnetization $\bm M_\mathrm{Fe}(\theta,\phi; T)$
and the total magnetization $\bm M(\theta,\phi;T)$.
}
\label{fig:angles}
\end{figure}

In this study,
the parameters we use for the $R_2$Fe$_{14}$B magnets were determined systematically by Yamada et al.\cite{Yamada1988}.
The temperature dependence of the saturated magnetization of the Y$_2$Fe$_{14}$B magnets was employed as $M_\mathrm{Fe}\mu_\mathrm{Fe}(T)$,
and its value at zero temperature is given by $M_\mathrm{Fe}=31.4$ $\mu_\mathrm{B}$/f.u.\cite{Hirosawa1986,Sagawa1987}.
To obtain its continuous values at nonzero temperature,
the Kuz'min formula has been widely used\cite{Kuzmin2005,Kuzmin2010,GomezEslava2016,Diop2016,Ozaki2017,Miura2018},
which is given as
\begin{align}
\mu_\mathrm{Fe}(T)=\left[1-s\left(\frac{T}{T_\mathrm{C}}\right)^{3/2}-(1-s)\left(\frac{T}{T_\mathrm{C}}\right)^p\right]^{1/3}.
\label{eq:mu}
\end{align}
Here, the Curie temperature $T_\mathrm{C}$ is defined for the target $R_2$Fe$_{14}$B magnet.
We confirmed\cite{Miura2018} that selecting the values of $s=1/2$ and $p=5/2$ for the shape parameters provides a good fit with the experimental result
of the Y$_2$Fe$_{14}$B magnet\cite{Hirosawa1986,Sagawa1987}.
Then, the total magnetization of the $R_2\mathrm{Fe}_{14}\mathrm{B}$ magnet is given by
\begin{align}
\bm M(\theta,\phi;T):=\bm M_\mathrm{Fe}(\theta,\phi;T)+\bm M_R(\theta,\phi;T),
\label{eq:mag}
\end{align}
where $\bm M_R(\theta,\phi;T)$ is the $R$ magnetization in units of [$\mu_\mathrm{B}$/f.u.] induced by the presence of the Fe magnetization and is defined by
\begin{subequations}
\begin{align}
M_R^x(\theta,\phi;T)&:=\frac{1}{2}\sum_{i=\mathrm{f},\mathrm{g}}\left[
m_{Ri}^x(\theta,\phi;T)
+
m_{Ri}^y(\theta,\phi+\pi/2;T)
\right],
\\
M_R^y(\theta,\phi;T)&:=\frac{1}{2}\sum_{i=\mathrm{f},\mathrm{g}}\left[
m_{Ri}^y(\theta,\phi;T)
-
m_{Ri}^x(\theta,\phi+\pi/2;T)
\right],
\\
M_R^z(\theta,\phi;T)&:=\frac{1}{2}\sum_{i=\mathrm{f},\mathrm{g}}\left[
m_{Ri}^z(\theta,\phi;T)
+
m_{Ri}^z(\theta,\phi+\pi/2;T)
\right].
\end{align}
\label{eq:magdef}
\end{subequations}
Here, we defined the magnetic moment of the $R$ ion on a site $i$ as
\begin{align}
\bm m_{Ri}(\theta,\phi;T):=-\Tr\E^{-\left[\mathcal{H}^i(\theta,\phi;T)-f_R^i(\theta,\phi;T)\right]/(k_\mathrm{B}T)}(\vec{\mathcal{L}}+2\vec{\mathcal{S}})
\end{align}
and the free energy of the $R$ ion as
\begin{align}
f_R^i(\theta,\phi;T):=-k_\mathrm{B}T\ln\Tr\E^{-\mathcal{H}^i(\theta,\phi;T)/(k_\mathrm{B}T)}.
\end{align}
By rotating $\bm M_\mathrm{Fe}(\theta,\phi;T)$ from the z-axis by hand,
the direction of $\bm M_R(\theta,\phi;T)$ deviates from $\bm M_\mathrm{Fe}(\theta,\phi;T)$ in the presence of the CEF;
to describe this non-collinearity,
we introduce the new symbols of $\Theta$ and $\Phi$ as the zenithal and azimuthal angles of the total magnetization, respectively, as shown in Fig. \ref{fig:angles}.

\subsection{Numerical analyses of Nd$_2$Fe$_{14}$B and Dy$_2$Fe$_{14}$B magnets across the entire temperature range}
\label{NA}

We explored the MA across the entire temperature range by computing the temperature-dependent magnetization and the temperature-dependent free energy density as functions of $\theta$ and $\phi$.
On the basis of our results, we show how the non-collinearity between the $R$ and Fe magnetizations appears
and how it effects the MA.

\begin{figure}[tb]
\centering
\includegraphics{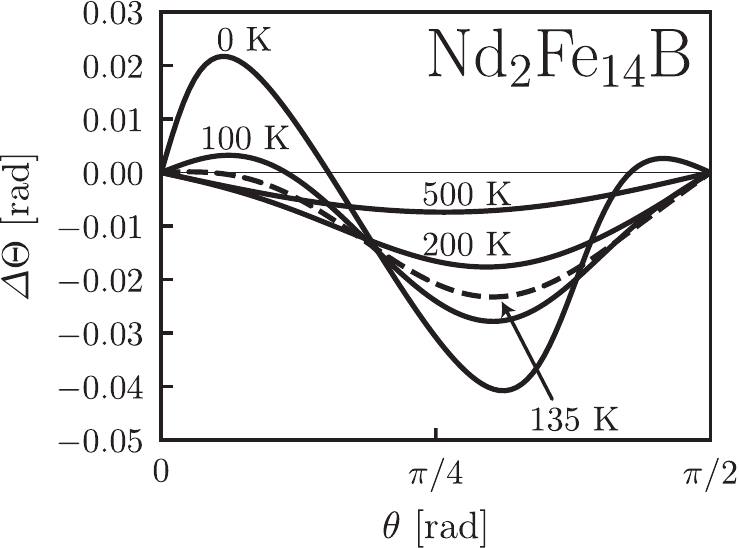}
\caption{%
Calculated angular difference, $\varDelta\Theta(\theta,T)$, between the total and Fe magnetizations as a function of the zenithal angle $\theta$ of the Fe magnetization in Nd$_2$Fe$_{14}$B compounds.
The number on each line denotes the temperature $T$,
and the dashed line is the result near the SRT temperature.
}
\label{fig:Nd_angles}
\end{figure}

First, we take Nd$_2$Fe$_{14}$B magnets
\footnote{$T_\mathrm{C}=586$ K for the Nd$_2$Fe$_{14}$B magnet\cite{Hirosawa1986,Sagawa1987}, and the parameters included in $\mathcal{H}^i(\theta,\phi;T)$ are listed
in Ref. \citenum{Yamada1988}.}
as an example of magnets, the NCE of which is small.
Figure \ref{fig:Nd_angles} shows the angular difference defined by
\begin{align}
\varDelta\Theta(\theta, T):=\Theta(\theta,\phi=0;T)-\theta,
\end{align}
as a function of $\theta$ for the compound Nd$_2$Fe$_{14}$B at several temperatures.
In the low-temperature range below the spin-reorientation transition (SRT) temperature ($T\sim 130$ K\cite{Yamada1988}),
these compounds exhibit complex behavior as shown by the lines for $T=0$, 100, and 135 K.
Above the SRT temperature,
we can observe that $\varDelta\Theta<0$ because $\bm M_\mathrm{Nd}(\theta,\phi;T)$ tends to naively orient along the $+$z-axis,
and that $\abs{\varDelta\Theta}$ monotonically decreases with increasing temperature.
In Nd$_2$Fe$_{14}$B magnets,
$\abs{\varDelta\Theta}$ has an extremely low value over the entire temperature range,
and thus we conclude that the NCE is negligibly small as assumed in our previous studies\cite{Sasaki2015,Miura2015,Miura2018}.
\begin{figure}[tb]
\centering
\includegraphics{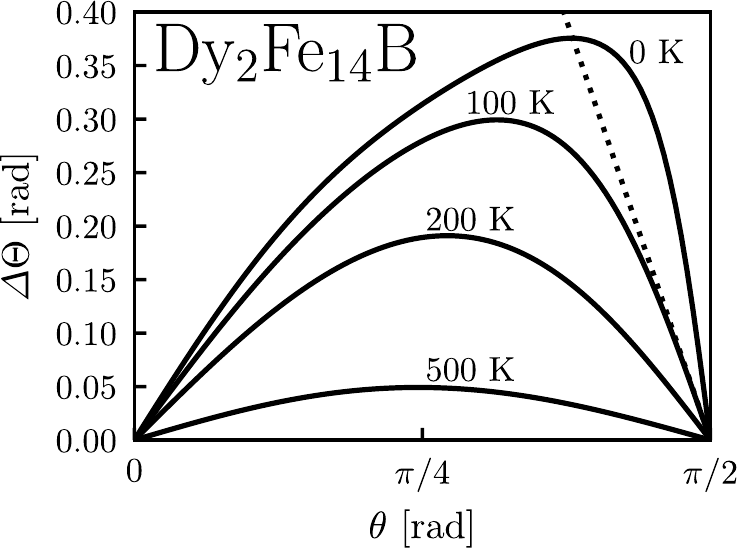}
\caption{%
Calculated angular difference, $\varDelta\Theta(\theta,T)$, between the total and Fe magnetizations as a function of the zenithal angle $\theta$ of the Fe magnetization in Dy$_2$Fe$_{14}$B compounds.
The number on each line denotes the temperature $T$,
and the dotted line represents $\pi/2-\theta$.
}
\label{fig:Dy_angles}
\end{figure}

In contrast,
the Dy$_2$Fe$_{14}$B magnets
\footnote{$T_\mathrm{C}=598$ K for the Dy$_2$Fe$_{14}$B magnet\cite{Hirosawa1986,Sagawa1987}, and the parameters included in $\mathcal{H}^i(\theta,\phi;T)$ are listed
in Ref. \citenum{Yamada1988}.}
exhibit high non-collinearity,
which is shown in Fig. \ref{fig:Dy_angles}.
Because Dy$_2$Fe$_{14}$B magnets do not have the SRT,
$\bm M_\mathrm{Dy}(\theta,\phi;T)$ tends to be oriented along the $-$z-axis over the entire temperature range; that is, $\varDelta\Theta>0$ for any temperature.
Here, let us focus on the intersection(s) of $\varDelta\Theta$ and the dotted line $\pi/2-\theta$ in Fig. \ref{fig:Dy_angles}.
At the intersection(s), $\Theta$ is equal to $\pi/2$.
In particular,
in the temperature range below approximately 100 K,
the intersection exists at an angle $\theta=\theta_\mathrm{o}<\pi/2$. %
That is, $\Theta$ overshoots $\pi/2$ at $\theta=\theta_\mathrm{o}$,
after which $\Theta>\pi/2$ for $\theta\in(\theta_\mathrm{o},\pi/2)$.
This fact becomes important when evaluating the MA of Dy$_2$Fe$_{14}$B magnets in the low-temperature range.
Here, it is shown that a large $\abs{\varDelta\Theta}$ remains even in the high-temperature range
compared with the Nd$_2$Fe$_{14}$B case.

\begin{figure}[tb]
\centering
\includegraphics[width=0.25\textwidth]{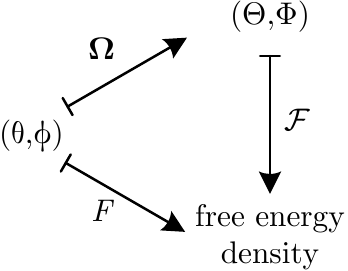}
\caption{%
Schematic view of the method used to obtain the map $\mathcal{F}$ from $(\Theta,\Phi)$ to the free energy density,
where the maps $\bm\Omega$ and $F$ can be calculated directly.
}
\label{fig:map}
\end{figure}

In accordance with the above results, we consider the NCE on the MA.
We define the total free energy density of $R_2$Fe$_{14}$B compounds as
\begin{align}
F(\theta,\phi;T):=F_R(\theta,\phi;T)+\kappa_\mathrm{Fe}(T)\sin^2\theta,
\label{eq:Ftheta}
\end{align}
where $F_R(\theta,\phi;T)$ is the contribution from the $R$ sublattice, which is given by 
\begin{align}
F_R(\theta,\phi;T):=\frac{2}{V_\mathrm{cell}}\sum_{i=\mathrm{f},\mathrm{g}}
\left[
f_R^i(\theta,\phi;T)
+
f_R^i(\theta,\phi+\pi/2;T)
\right]
,
\end{align}
where $V_\mathrm{cell}$ is the volume of the unit cell.
The second term in Eq. (\ref{eq:Ftheta}) is the contribution from the Fe sublattice,
where we use the first-order MAC of the Y$_2$Fe$_{14}$B magnet as $\kappa_\mathrm{Fe}(T)$,
which is expressed by a fitting form as\cite{Miura2018}
\begin{align}
\kappa_\mathrm{Fe}(T)
&=
\kappa_1^\mathrm{Fe}\mu_\mathrm{Fe}(T)^3\notag\\
&+
\frac{8}{7}\kappa_2^\mathrm{Fe}\left[\mu_\mathrm{Fe}(T)^3-\mu_\mathrm{Fe}(T)^{10}\right]\notag\\
&+
\frac{8}{7}\kappa_3^\mathrm{Fe}\left[\mu_\mathrm{Fe}(T)^3-\frac{18}{11}\mu_\mathrm{Fe}(T)^{10}+\frac{7}{11}\mu_\mathrm{Fe}(T)^{21}\right],
\end{align}
where the fitted parameters are given by
$\kappa_1^\mathrm{Fe}=0.77$ MJ/m$^3$,
$\kappa_2^\mathrm{Fe}=1.21$ MJ/m$^3$,
and
$\kappa_3^\mathrm{Fe}=0.11$ MJ/m$^3$.
To determine the NCE on the MA in $R_2$Fe$_{14}$B magnets,
we compute the total free energy density as a function of $\theta$ and $\phi$,
and next obtain this energy density as a function of $\Theta$ and $\Phi$ by the following process (see Fig. \ref{fig:map}):
(i) calculate $F(\theta,\phi;T)$ as a function of $(\theta,\phi)$ in Eq. (\ref{eq:Ftheta});
(ii) calculate $\Theta$ and $\Phi$ as a function of $(\theta,\phi)$;
(iii) regard $F(\theta,\phi;T)$ as $\mathcal{F}(\Theta,\Phi;T)$,
by noting that
if the values $(\theta_i,\phi_i)$ exist
such that $(\Theta,\Phi)=\bm\Omega(\theta_1,\phi_1)=\bm\Omega(\theta_2,\phi_2)=\cdots$,
then we put $\mathcal{F}(\Theta,\Phi;T)=\min[ F(\theta_1,\phi_1;T),F(\theta_2,\phi_2;T),\ldots]$,
where $\bm\Omega$ is the map from $(\theta,\phi)$ to $(\Theta,\Phi)$.

\begin{figure}[tbp]
\centering
\includegraphics[width=0.4\textwidth]{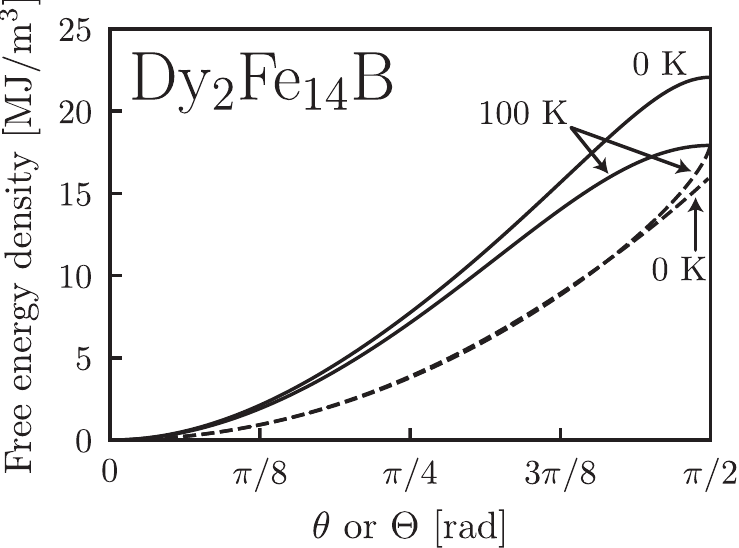}
\\
\includegraphics[width=0.4\textwidth]{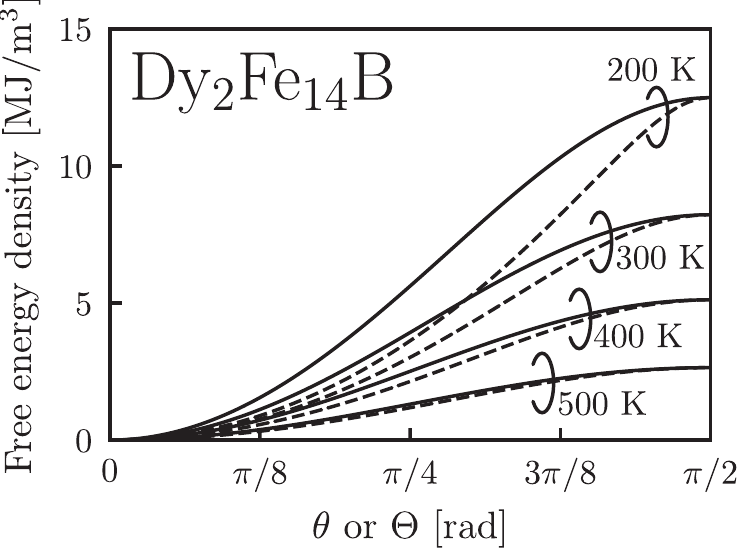}
\caption{Calculated zenithal angle dependence of the free energy density of the Dy$_2$Fe$_{14}$B compounds at each temperature.
The solid and dashed lines represent $F(\theta,\phi;T)$
and $\mathcal{F}(\Theta,\Phi;T)$, respectively,
at $\phi=0$.
}
\label{fig:F-Theta-Dy}
\end{figure}

The calculated angle dependence of the total free energy density of the compound Dy$_2$Fe$_{14}$B is shown in Fig. \ref{fig:F-Theta-Dy},
where the solid and dashed lines are the $\theta$ and $\Theta$ dependencies, respectively.
Across the entire temperature range, the stabilization angle is $\theta=0$;
the Dy magnetization tends to be oriented along the $-$z axis; hence, $\Theta\ge\theta$ in $0\le\theta\le\pi/2$ as shown in Fig. \ref{fig:Dy_angles}.
In the low-temperature range below approximately 100 K,
Fig. \ref{fig:Dy_angles} indicates that $\Theta>\pi/2$ when $\theta$ varies from $\theta_0$ to $\pi/2$,
and therefore, it is clear from Fig. \ref{fig:F-Theta-Dy} (upper) that the stabilization energy is lower than the fictitious stabilization energy estimated from the dependence on $\theta$.
Moreover, the dashed lines in Fig. \ref{fig:F-Theta-Dy} (upper) almost completely overlap,
which suggests that Dy$_2$Fe$_{14}$B compounds have a magnetic anisotropy that is robust against a rise in temperature.
In contrast,
as shown in Fig. \ref{fig:F-Theta-Dy} (lower),
the stabilization energy estimated from the $\Theta$ dependence (dashed lines) is equal to the fictitious stabilization energy from the $\theta$ dependence (solid lines) in the high-temperature range.
However, the initial rise in the $\theta$ dependence is clearly larger than that of the $\Theta$ dependence,
and therefore, the MACs would be overestimated if the estimation were to be based on the $\Theta$ dependence.

Lastly, in this section,
let us consider the MACs derived from $\varDelta\mathcal{F}(\Theta,\Phi;T):=\mathcal{F}(\Theta,\Phi;T)-\mathcal{F}(0,0;T)$ in Dy$_2$Fe$_{14}$B magnets.
We introduce a third-order fitting function for $\varDelta\mathcal{F}(\Theta,\Phi;T)$ as
\begin{align}
\varDelta\mathcal{F}_\mathrm{fit}(\Theta,\Phi;T)
&:=K_1^\mathrm{fit}(T)\sin^2\Theta\notag\\
&+K_2^\mathrm{fit}(T)\sin^4\Theta+K_3^\mathrm{fit}(T)\sin^6\Theta,
\label{eq:fit}
\end{align}
where $K_n^\mathrm{fit}(T)$ is the $n$th-order fitted MACs,
and we have ignored the cases in which Dy$_2$Fe$_{14}$B depends on $\Phi$ because this dependence is sufficiently small.
Here, note that we have performed the fitting calculations in a $\Theta$ range near $\Theta\sim 0$ (corresponding to $\theta\in[0,\pi/20]$)
because the fitting form in Eq. (\ref{eq:fit}) is clearly not appropriate
for the singular shape of the dashed lines near $\Theta=\pi/2$ in Fig. \ref{fig:F-Theta-Dy} (upper).
\begin{figure}[tb]
\centering
\includegraphics{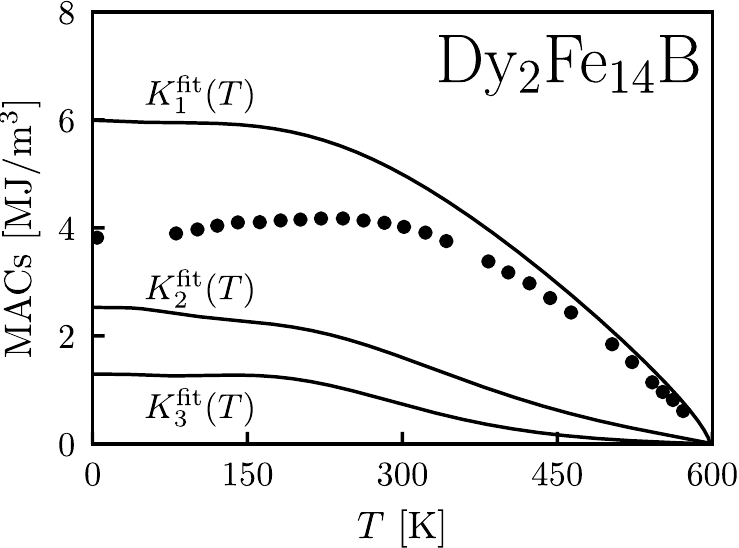}
\caption{MACs of the Dy$_2$Fe$_{14}$B compounds as a function of the temperature $T$.
The solid lines are the results calculated as part of this study,
and the solid circles are the experimental results of the first-order MAC\cite{Hirosawa1985}.
}
\label{fig:DyK-T}
\end{figure}
Figure \ref{fig:DyK-T} shows the values obtained for $K_n^\mathrm{fit}(T)$ by fitting $\varDelta\mathcal{F}_\mathrm{fit}(\Theta,\Phi;T)$ to $\varDelta\mathcal{F}(\Theta,\Phi;T)$.
It is noteworthy that the broad plateau in the low-temperature range is reproduced,
although the calculated $K_1^\mathrm{fit}(T)$ is larger than the experimental results by approximately 2 MJ/m$^3$;
for quantitative comparison,
we notice that the experimental first-order MAC has been estimated from experimental anisotropy fields assuming the absence of higher-order MACs.
This indicates that this broad plateau originates from the robustness of the MA against a rise in temperatures as mentioned in the previous paragraph.
That is, the presence of the plateau reflects the effects of the non-collinearity;
in fact, under the assumption of collinear magnetizations, such a broad plateau is not obtained for Dy$_2$Fe$_{14}$B magnets as demonstrated by Ito et al.\cite{Ito2016}.
Furthermore,
no less important is that the damping of $K_2^\mathrm{fit}(T)$ and $K_3^\mathrm{fit}(T)$ is slow in the high-temperature range.
The slow damping of the higher-order MACs can also be understood in terms of the NCE, which is considered in the next section.
To conclude this section,
we emphasize that the non-collinearity of Dy$_2$Fe$_{14}$B magnets is not negligible, especially when evaluating the MA.

\subsection{Perturbative expressions for MACs with NCEs in the high-temperature range}
\label{PE}

The importance of considering the effect of the non-collinearity between the Dy and Fe magnetizations is explained in the previous section.
In this section, we first derive explicit microscopic expressions for the MA
by taking into account the NCE in the high-temperature range,
and subsequently apply the result to Dy$_2$Fe$_{14}$B and other $R_2$Fe$_{14}$B magnets.
Here, we consider only the ground $J$ multiplet and ignore the $J$-mixing effects.
Although some light $R$ ions such as Pr, Nd, and Sm exhibit a large $J$-mixing effect on MA
as pointed out by several authors\cite{Yamada1988,Kuzmin1994,Magnani2000,Magnani2001,Kuzmin2002,Magnani2003},
we determined the order in which the non-collinearity can be ignored to be
$0\le 2\abs{\varDelta\Theta_\mathrm{Nd}}/\pi < 0.03$ as is evident from the previous section,
and $0\le 2\abs{\varDelta\Theta_\mathrm{Pr}}/\pi < 0.04$ and $0\le 2\abs{\varDelta\Theta_\mathrm{Sm}}/\pi < 0.01$
by further numerical calculations with finite SOI.
Thus, we discuss the NCE only for heavy $R$ elements by using the CEF and EXF parameters reported by Yamada et al.\cite{Yamada1988}.

The 4f total Hamiltonian at a site $i$ within the ground multiplet $J$ can be expressed in terms of the total angular momentum operator of the 4f electrons, $\vec{\mathcal{J}}$,
on the basis of the Wigner-Eckart theorem, and $\vec{\mathcal{J}}^2$ is the constant $J(J+1)$:
\begin{subequations}
\begin{align}
&\mathcal{V}_\mathrm{CEF}^i\to\bar{\mathcal{V}}_\mathrm{CEF}^i:=\sum_{(\ell,m)}B_\ell^m(i)\mathcal{O}_\ell^m(\vec{\mathcal{J}}),\\
&\lambda\vec{\mathcal{L}}\cdot\vec{\mathcal{S}}\to\frac{\lambda}{2}\left[J(J+1)-\vec{\mathcal{L}}^2-\vec{\mathcal{S}}^2\right]
=\mathrm{const.},\\
&-2\vec{\mathcal{S}}\cdot \bm H_\mathrm{EXF}(\theta,\phi;T) \to -2(g-1)\vec{\mathcal{J}}\cdot\bm H_\mathrm{EXF}(\theta,\phi;T),
\end{align}
\end{subequations}
that is,
\begin{align}
\mathcal{H}^i(\theta,\phi;T)&\to
\bar{\mathcal{V}}_\mathrm{CEF}^i-2(g-1)\vec{\mathcal{J}}\cdot\bm H_\mathrm{EXF}(\theta,\phi;T)+\mathrm{const.},
\end{align}
where
$\mathcal{O}_\ell^m(\vec{\mathcal{J}})$ is the Stevens operator\cite{Stevens1952},
and the range of the $(\ell,m)$ summation is limited to $(\ell,m)=(2,0)$, $(2,-2)$, $(4,0)$, $(4,-2)$, $(4,4)$, $(6,0)$, $(6,-2)$, $(6,4)$, and $(6,-6)$
by the symmetry of the CEF.
On the basis of the definitions Eqs. (\ref{eq:EXF2}) and (\ref{eq:mu}), we can perform the perturbative expansion for the free energy density of the $R$ ions, $F_R(\theta,\phi;T)$, with respect to the dimensionless parameter $\mu_\mathrm{Fe}(T)$ in the high-temperature range.
In this expansion, $F_R(\theta,\phi;T)$ only has even powers of $\mu_\mathrm{Fe}(T)$ owing to the time inversion symmetry,
and thus, the lowest contribution arises from the second-order of $\mu_\mathrm{Fe}(T)$ as
\begin{align}
\varDelta F_R(\theta,\phi;T):=
F_R(\theta,\phi;T)-F_R(0,0;T)
=
\kappa_R(T)
\sin^2\theta
+
O(\mu_\mathrm{Fe}(T)^4)
,
\end{align}
where
\begin{align}
\kappa_R(T)
:=
\frac{8}{V_\mathrm{cell}}
\frac{\left[2(g-1)H_\mathrm{EXF} \right]^2}{2}
\mu_\mathrm{Fe}(T)^2
\left[
\chi_z(T)-\chi_x(T)
\right]
,
\label{eq:kr}
\end{align}
and the factor $8/V_\mathrm{cell}$ represents the concentration of the $R$ ions.
Furthermore, we introduced
\begin{align}
\chi_\alpha(T):=\frac{1}{2}\sum_{i=\mathrm{f},\mathrm{g}}\int_0^{(k_\mathrm{B}T)^{-1}}\D\tau \Braket{\E^{\tau\bar{\mathcal{V}}_\mathrm{CEF}^i}
\mathcal{J}_\alpha
\E^{-\tau\bar{\mathcal{V}}_\mathrm{CEF}^i}
\mathcal{J}_\alpha}_T,
\end{align}
where
$\braket{\cdots}_T$ denotes the statistical average in $\bar{\mathcal{V}}_\mathrm{CEF}^i$ at a temperature $T$,
and further, we used the symmetry $\chi_x(T)=\chi_y(T)$ in the derivation of Eq. (\ref{eq:kr}).
Thus, the total free energy density is given by
\begin{align}
\varDelta F(\theta,\phi;T):=
F(\theta,\phi;T)-F(0,0;T)
=
\kappa_1(T)\sin^2\theta
+
O(\mu_\mathrm{Fe}(T)^4)
,
\label{eq:F}
\end{align}
where
\begin{align}
\kappa_1(T):=\kappa_\mathrm{Fe}(T)+\kappa_R(T).
\label{eq:kappa1}
\end{align}
If one considers a magnet with a low non-collinearity between the $R$ and Fe moments such as a Nd$_2$Fe$_{14}$B magnet,
then it becomes possible to conclude that $K_1(T)\simeq\kappa_1(T)$.
However, as we have mentioned, this assumption is not always satisfied.

We describe the total magnetization within the same framework with the aim of taking the NCE into account.
By perturbatively expanding Eq. (\ref{eq:magdef}) with respect to $\mu_\mathrm{Fe}(T)$ again,
$\bm M_R(\theta,\phi;T)$ only has odd powers of $\mu_\mathrm{Fe}(T)$,
and we obtain
\begin{align}
M_R^\alpha(\theta,\phi;T)
&=
-4g(g-1)
H_\mathrm{EXF} 
\mu_\mathrm{Fe}(T)
m_\mathrm{Fe}^\alpha(\theta,\phi)
\chi_\alpha(T)
+
O(\mu_\mathrm{Fe}(T)^3)
.
\end{align}
Then, we determine the relationship between the directions of $\bm M(\theta,\phi;T)$ and $\bm M_\mathrm{Fe}(\theta,\phi;T)$ as
\begin{align}
\sin^2\theta=
\frac{
[1+\delta(T)]\sin^2\Theta
}{1+\delta(T)\sin^2\Theta}
+
O(\mu_\mathrm{Fe}(T)^2)
,
\label{eq:sin}
\end{align}
where we defined the non-collinearity factor as
\begin{align}
\delta(T)
&:=
\frac{-8g(g-1)H_\mathrm{EXF} \left\{M_\mathrm{Fe} -2g(g-1)H_\mathrm{EXF} [\chi_z(T)+\chi_x(T)]\right\}}{
\left[
M_\mathrm{Fe} -4g(g-1)H_\mathrm{EXF} \chi_x(T)
\right]^2
}
\left[
\chi_z(T)-\chi_x(T)
\right]
\label{eq:delta}
,
\end{align}
where we notice that $\delta(T_\mathrm{C})$ does not vanish.
Substituting Eq. (\ref{eq:sin}) into Eq. (\ref{eq:F}) and assuming $\abs{\delta(T)}<1$,
the MACs, including the NCE up to the second-order $\mu_\mathrm{Fe}(T)$, are expressed as
\begin{align}
K_n(T)=\kappa_1(T)[1+\delta(T)][-\delta(T)]^{n-1}
\quad\mathrm{for}\quad n\ge 1.
\end{align}
Because the effect of the CEF on both $\kappa_R(T)$ and $\delta(T)$ is reflected through $\chi_\alpha(T)$,
let us try to expand $\chi_\alpha(T)$ with respect to an expansion parameter $y:=B_l^m/(k_\mathrm{B}T)$
to determine the relation between the MACs and the CEF:
\begin{align}
\chi_\alpha(T)=\sum_{n=0}^\infty\frac{x_\alpha^{(n)}}{(k_\mathrm{B}T)^{n+1}}
=
\frac{x_\alpha^{(0)}}{k_\mathrm{B}T}
+
\frac{x_\alpha^{(1)}}{(k_\mathrm{B}T)^2}
+
\frac{x_\alpha^{(2)}}{(k_\mathrm{B}T)^3}
+
\frac{O(y^3)}{k_\mathrm{B}T}
,
\label{eq:chi}
\end{align}
where the temperature-independent coefficients are given by
\begin{subequations}
\begin{align}
x_\alpha^{(0)}&=\frac{J(J+1)}{3},\\
x_\alpha^{(1)}&=-\frac{1}{2}\sum_{i=\mathrm{f},\mathrm{g}}
\frac{\mathrm{Tr\ }
\bar{\mathcal{V}}_\mathrm{CEF}^i(\mathcal{J}_\alpha)^2
}{2J+1}
,\\
x_\alpha^{(2)}&=\frac{1}{2}\sum_{i=\mathrm{f},\mathrm{g}}
\frac{\mathrm{Tr}
\left[
2(\bar{\mathcal{V}}_\mathrm{CEF}^i)^2(\mathcal{J}_\alpha)^2
+
(\bar{\mathcal{V}}_\mathrm{CEF}^i\mathcal{J}_\alpha)^2
-
(\bar{\mathcal{V}}_\mathrm{CEF}^i)^2\vec{\mathcal{J}}^2
\right]
}{6(2J+1)},
\\
&\cdots.\notag
\end{align}
\end{subequations}
Substituting Eq. (\ref{eq:chi}) into Eqs. (\ref{eq:kr}) and (\ref{eq:delta}),
we obtain
\begin{align}
\kappa_R(T)
&=
\kappa_R^{(1)}(T)
\left(
1+
\frac{\zeta}{T}
+O\left(
y^2
\right)
\right)
,
\label{eq:krexpand}
\\
\delta(T)
&=
\delta^{(1)}(T)
\left[
1
+
O(y)
\right],
\end{align}
where
\begin{align}
\kappa_R^{(1)}(T)
&:=
\frac{4}{V_\mathrm{cell}}
\left[2(g-1)H_\mathrm{EXF} \right]^2
\mu_\mathrm{Fe}(T)^2
\frac{
x_z^{(1)}-x_x^{(1)}
}{(k_\mathrm{B}T)^2}
\\
\delta^{(1)}(T)
&:=
\frac{-8g(g-1)H_\mathrm{EXF} 
}{
M_\mathrm{Fe}-4g(g-1)H_\mathrm{EXF} 
J(J+1)/(3 k_\mathrm{B}T)
}
\frac{
x_z^{(1)}
-
x_x^{(1)}
}{(k_\mathrm{B}T)^2}
,
\end{align}
and
\begin{align}
\zeta
&:=
\frac{1}{k_\mathrm{B}}
\frac{
x_z^{(2)}-x_x^{(2)}
}{x_z^{(1)}-x_x^{(1)}}.
\end{align}
If the calculation of $K_n(T)$ takes into consideration $\delta^{(1)}(T)$, which describes the NCE in the leading order,
then there is no reason to ignore a correction from $\zeta/T$ in Eq. (\ref{eq:krexpand}) in general cases,
because both are of the same order of $y$.
However, if the targets are limited to $R_2$Fe$_{14}$B magnets,
we can conclude that the $\zeta/T$ term is negligible in the high-temperature range as in Table \ref{tab:zeta}.
As a result,
ignoring $\zeta$ in the assumption of $\zeta/T\ll 1$
allows us to estimate the MACs, up to the second-order $\mu_\mathrm{Fe}(T)$, by using
\begin{subequations}
\begin{align}
K_n(T)&\simeq
\left[
\kappa_\mathrm{Fe}(T)
+
\kappa_R^{(1)}(T)
\right]
\left[
1+\delta^{(1)}(T)
\right]
\left[
-\delta^{(1)}(T)
\right]^{n-1}
\quad\mathrm{for}\quad n\ge 1,
\label{eq:Knre}
\intertext{and explicit forms are given as}
\kappa_R^{(1)}(T)&=
-\frac{4(g-1)^2
J(J+1)(2J-1)(2J+3)
}{5V_\mathrm{cell}}
\left(\frac{\mu_\mathrm{Fe}(T)
H_\mathrm{EXF} 
}{k_\mathrm{B}T}\right)^2
\sum_{i=\mathrm{f},\mathrm{g}}
B_2^0(i)
,
\label{eq:KRre}\\
\delta^{(1)}(T)
&=
\frac{
2g(g-1)J(J+1)(2J-1)(2J+3)H_\mathrm{EXF} 
}{
5\left[
M_\mathrm{Fe} 
-4 g(g-1)
J(J+1)
H_\mathrm{EXF} 
/(3k_\mathrm{B}T)
\right]
(k_\mathrm{B}T)^2
}
\sum_{i=\mathrm{f},\mathrm{g}}
B_2^0(i)
.
\label{eq:NCFre}
\end{align}
\label{eq:re}%
\end{subequations}
\begin{table}[tb]
  \centering
  \caption{Calculated values of $\zeta$ in units of [K].}
    \begin{tabular}{ccccccccc}
    \hline
     Tb    & Dy    & Ho    & Er    & Tm    & Yb \bigstrut\\
    \hline
    $\quad13.24\quad$ & $\quad-3.52\quad$ & $\quad-1.50\quad$  & $\quad1.54\quad$  & $\quad3.45\quad$  & $\quad-12.56\quad$ \bigstrut\\
    \hline
    \end{tabular}%
  \label{tab:zeta}%
\end{table}%
The main results of this study are expressed by Eq. (\ref{eq:re}).
Near the Curie temperature, the MA exhibits explicit temperature dependence:
$\mu_\mathrm{Fe}(T)/T\simeq 2^{1/3}(1-T/T_\mathrm{C})^{1/3}/T_\mathrm{C}$,
$\delta^{(1)}(T)\simeq\delta^{(1)}(T_\mathrm{C})$,
and $\kappa_\mathrm{Fe}(T)/\kappa_R^{(1)}(T)\ll 1$;
thus, the temperature dependence of the MACs is proportional to $(1-T/T_\mathrm{C})^{2/3}$.
The high-temperature expansion [Eq. (\ref{eq:KRre})] was first derived by Kuz'min\cite{Kuzmin1995}.
If the NCEs are ignorable, i.e., $\abs{\delta(T)}\ll 1$,
then, on the basis of Eq. (\ref{eq:re}), it can be confirmed that $K_1(T)\simeq \kappa_1^{(1)}(T)$ and the higher-order MACs are negligible at high temperatures.
In this sense, we have naturally extended Kuz'min's result in consideration of the NCEs.

\begin{figure}[tb]
\centering
\includegraphics[width=0.4\textwidth]{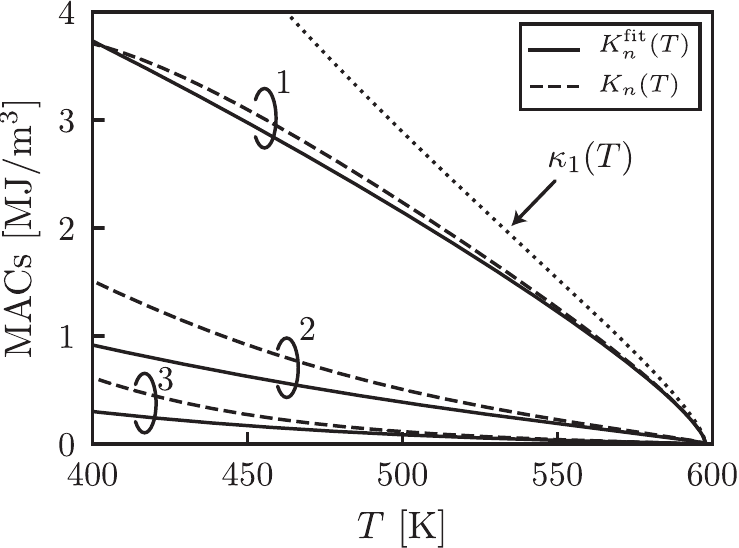}
\caption{%
Comparison of the MACs calculated by the fitting method (solid lines) with those obtained by the high-temperature expansion (dashed lines)
in the Dy$_2$Fe$_{14}$B magnet; the solid lines represent the same result as in Fig. \ref{fig:DyK-T}.
The number on each ring denotes the value of $n$.
}
\label{fig:Dy_MACwithNCF}
\end{figure}

Lastly, in this section, let us compare the present results with the fitted MACs.
The case for Dy$_2$Fe$_{14}$B magnets is shown in Fig. \ref{fig:Dy_MACwithNCF}, where
the solid lines are $K_1^\mathrm{fit}(T)$, $K_2^\mathrm{fit}(T)$, and $K_3^\mathrm{fit}(T)$, the same as in Fig. \ref{fig:DyK-T},
and the dashed lines are $K_1(T)$, $K_2(T)$, and $K_3(T)$ by using Eq. (\ref{eq:re}) with 
$g=4/3$,
$J=15/2$,
$B_2^0(\mathrm{f})/k_\mathrm{B}=B_2^0(\mathrm{g})/k_\mathrm{B}=-1.392$ K,
and
$H_\mathrm{EXF} /k_\mathrm{B}=145$ K\cite{Yamada1988}.
Then, we can observe that the high-temperature expansion provides a good approximation for the solid lines.
As mentioned in Sect. \ref{NA},
the NCE in Nd$_2$Fe$_{14}$B magnets is low but high in Dy$_2$Fe$_{14}$B magnets.
If the first-order MAC of Dy$_2$Fe$_{14}$B magnets is evaluated without the NCE,
then it is overestimated by approximately 20\% at 500 K as illustrated by $\kappa_1(T)$ in Eq. (\ref{eq:kappa1}) with Eq. (\ref{eq:KRre}).
Therefore, 
the \textit{decomposition} of $\kappa_1(T)$ into $K_1(T)$, $K_2(T)$, $\ldots$ by the NCE, as expressed by Eq. (\ref{eq:Knre}),
is continued up to the Curie temperature because $\delta(T_\mathrm{C})$ does not vanish,
and this is the reason that $K_2(T)$ survives even near the Curie temperature.
As a consequence,
we can understand that
the non-collinearity arises from the non-negligible $K_2(T)$ of Dy$_2$Fe$_{14}$B magnets at high temperatures, as mentioned in Sect. \ref{NA},
and also from $B_2^0(i)$ and $H_\mathrm{EXF}$.
In contrast, in a small non-collinearity system,
$K_2(T)$ is mainly induced by $B_4^0(i)$ and/or $B_6^0(i)$\cite{Kuzmin1995};
thus, the mechanism essentially differs.

\begin{figure}[tb]
\centering
\includegraphics{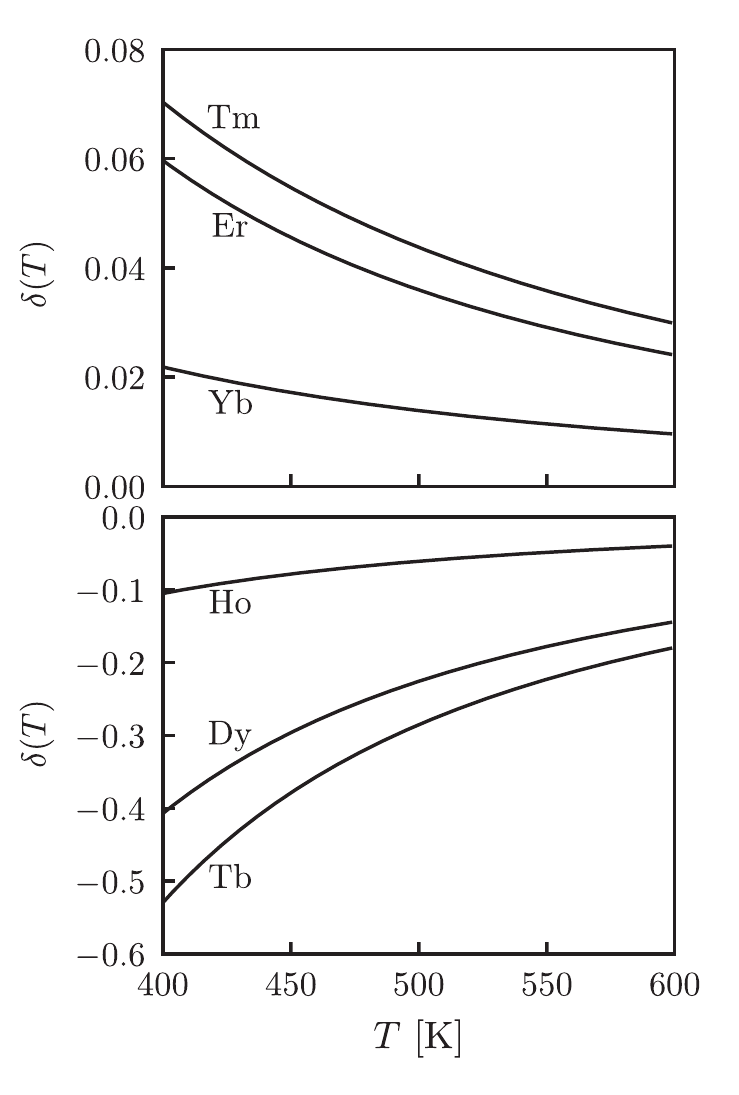}
\caption{%
Temperature dependence of the non-collinearity factor $\delta(T)$ as a function of temperature $T$ calculated using Eq. (\ref{eq:NCFre}),
for $R_2$Fe$_{14}$B magnets
($R=$Tm, Er, Yb, Ho, Dy, and Tb).
}
\label{fig:NCF}
\end{figure}

\subsection{Practical expressions for MACs in rare-earth intermetallics}
The simple expression at zero temperature [Eq. (\ref{eq:YKYN0})],
\begin{align}
K_1^R(0)=-3f_2 B_2^0-40 f_4 B_4^0 -168 f_6 B_6^0,
\end{align}
motivated the evaluation of $B_2^0$, $B_4^0$, and $B_6^0$, especially from first principles\cite{Miyake2018}.
However, as is well known,
the temperature dependence of the MA of RE magnets is complex,
and thus, this expression is inappropriate to evaluate the MA in the high-temperature range, which is important in practical situations used in electric vehicle motors.
Here, for the reader's convenience,
we provide a useful form of the expressions obtained in the previous section
to allow us to immediately estimate the temperature-dependent MA for two sublattice RE magnets consisting of RE and transition elements.

Although the three CEF coefficients are needed to evaluate the zero-temperature MA because they have the same order,
we do not exert effort to evaluate the higher-order CEF coefficients in the high-temperature range.
This is because the high-temperature MA is dominated only by $B_2^0$ as explained in the previous section.
Now, when
one has
a CEF coefficient, $\bar B_2^0$ [K], which is the average value of $B_2^0$ over the total RE ions,
and
an EXF, $H_\mathrm{RE}$ [K], by which the effective exchange energy is represented as $-2(g-1)\mathcal{J}_z H_\mathrm{RE}$ at zero temperature,
the $n$th-order MACs in the high-temperature range can be estimated as
\begin{align}
K_n^\mathrm{Total}(T)=
\left[
\kappa_\mathrm{T}(T)
+
\kappa_\mathrm{RE}(T)
\right]
\left[
1+\delta_\mathrm{RE-T}(T)
\right]
\left[
-\delta_\mathrm{RE-T}(T)
\right]^{n-1}
,
\end{align}
in units of [K/$V_\mathrm{cell}$],
where
$\kappa_\mathrm{T}(T)$ is the experimental first-order MAC within the transition metal sublattice,
and
\begin{align}
\kappa_\mathrm{RE}(T)&=-A n_\mathrm{RE} \left(\frac{H_\mathrm{RE}\mu_\mathrm{T}(T)}{T}\right)^2\bar B_2^0,\\
\delta_\mathrm{RE-T}(T)&=\frac{B n_\mathrm{RE} H_\mathrm{RE} \bar B_2^0}{T(T M_\mathrm{T} -C n_\mathrm{RE} H_\mathrm{RE})},
\end{align}
where
$n_\mathrm{RE}$ is the number of RE ions in the unit cell,
$M_\mathrm{T}$ [$\mu_\mathrm{B}/V_\mathrm{cell}$] is the saturated magnetization of the transition metal sublattice at zero temperature,
and $A$, $B$, and $C$ are geometric coefficients determined by $g$ and $J$ of each of the rare-earth elements listed in Table \ref{tab:abc}.
The definitions are immediately obtained from Kuz'min's result [Eq. (\ref{eq:KRre})] and the present result [Eq. (\ref{eq:NCFre})].
$\mu_\mathrm{T}(T)$ can be expressed by the Kuz'min formula\cite{Kuzmin2005} as
\begin{align}
\mu_\mathrm{T}(T)=
\left[
1-s\left(\frac{T}{T_\mathrm{C}}\right)^{3/2}
-
(1-s)
\left(\frac{T}{T_\mathrm{C}}\right)^p
\right]^{1/3},
\end{align}
where $s$ and $p$ are previously reported shape parameters\cite{Kuzmin2005,Kuzmin2010,Miura2018}.
For example, $\delta(T)$ for Dy$_2$Fe$_{14}$B magnets in Fig. \ref{fig:NCF} can be reproduced
by setting
$n_\mathrm{RE}=8$,
$H_\mathrm{RE}=145$ K,
$\bar B_2^0=-1.392$ K,
$M_\mathrm{T}=31.4\times 4\ \mu_\mathrm{B}/V_\mathrm{cell}$,
$B=2856$,
$C=170/9$,
$s=1/2$,
$p=5/2$,
and $T_\mathrm{C}=598$ K.
In addition to Dy, the calculated non-collinearity factors of other magnets are shown in Fig. \ref{fig:NCF}.
The values of the CEF and EXF parameters are those of Yamada et al.\cite{Yamada1988}.
The sign of $\delta(T)$ is equal to $(g-1)\sum_{i=\mathrm{f},\mathrm{g}} B_2^0(i)$ as shown in Eq. (\ref{eq:NCFre}),
and the sign of $B_2^0(i)$ is equal to $\theta_2 A_2^0(i)$, where $\theta_2$ is the Stevens factor and $A_2^0(i)$ is the CEF parameter.
Because $A_2^0(i)>0$ for $R_2$Fe$_{14}$B magnets,
the $R$ dependence of the sign of $\delta(T)$ is determined by $(g-1)\theta_2$,
in which
$g-1>0$ for the heavy $R$ ions,
and $\theta_2>0$ for Er, Tm, and Yb,
and $\theta_2<0$ for Tb, Dy, and Ho.
We can observe that Dy and Tb especially exhibit a large NCE.

Note that the present results do not include the effects of $J$-mixing.
In general, elements from the light RE series have a larger $J$-mixing effect than the heavy ones\cite{Yamada1988,Kuzmin1994,Kuzmin2002,Magnani2003}.
Whether the $J$-mixing effect becomes serious for the NCE in general RE intermetallics is not a trivial matter,
although we were able to ignore the NCE for $R=$ Pr, Nd, and Sm in the case of $R_2$Fe$_{14}$B.
For several light $R$,
the $J$-mixing effect on $\kappa_\mathrm{RE}(T)$ cannot be ignored, especially for Sm.
This problem was detailed by Kuz'min\cite{Kuzmin2002} and Magnani et al.\cite{Magnani2003},
and explicit expressions for $\kappa_\mathrm{RE}(T)$ were provided.

\begin{table}[htbp]
  \centering
  \caption{Calculated geometric coefficients for each RE ion}
    \begin{tabular}{ccccc}
    \hline
    RE ion       & $A$  & $B$  & $C$ \bigstrut\\
    \hline
    Ce$^{3+}$       & $8/7$  & $-96/7$ & $-5/7$ \bigstrut[t]\\
    Pr$^{3+}$       & $308/25$ & $-2464/25$ & $-32/15$ \\
    Nd$^{3+}$       & $1944/55$ & $-10368/55$ & $-36/11$ \\
    Pm$^{3+}$       & $1232/25$ & $-3696/25$ & $-16/5$ \\
    Sm$^{3+}$       & $200/7$ & $-160/7$ & $-25/21$ \\
    Gd$^{3+}$       & $189$  & $756$ & $21$ \\
    Tb$^{3+}$       & $693/2$ & $2079$ & $21$ \\
    Dy$^{3+}$       & $357$  & $2856$ & $170/9$ \\
    Ho$^{3+}$      & $513/2$ & $2565$ & $15$ \\
    Er$^{3+}$      & $3213/25$ & $38556/25$ & $51/5$ \\
    Tm$^{3+}$      & $77/2$ & $539$  & $49/9$ \\
    Yb$^{3+}$      & $27/7$ & $432/7$ & $12/7$ \\
    \hline
    \end{tabular}%
  \label{tab:abc}%
\end{table}%

\section{Summary}
\label{summary}

We showed that Dy$_2$Fe$_{14}$B magnets have a large NCE on the MA compared with Nd$_2$Fe$_{14}$B magnets,
and that the NCE in Dy$_2$Fe$_{14}$B magnets yields a plateau of $K_1(T)$ in the low-temperature range
and a non-negligible $K_2(T)$ in the high-temperature range.
Furthermore, we derived microscopic expressions [Eq. (\ref{eq:re})] for $K_n(T)$ with NCEs by using the high-temperature expansion,
and showed that these expressions were in a form extending Kuz'min's collinear result [Eq. (\ref{eq:KRre})].
In homogeneous local moment systems,
$B_4^0$ and $B_6^0$ are important for the rise of $K_2(T)$,
and $K_2(T)$ rapidly decays with increasing temperature as represented by Eq. (\ref{eq:K2CEF}).
However, interestingly, in high non-collinear system, $K_2(T)$ survives even in the high-temperature range
because of the presence of $B_2^0$ as given in Eq. (\ref{eq:re}).

In terms of Eqs. (\ref{eq:KRre}) and (\ref{eq:NCFre}),
the main contribution to the MA comes from both $B_2^0(s)$ and $H_\mathrm{EXF}$ in the high-temperature range,
whereas the higher-order CEF parameters are not effective.
This is also an interesting result for the field of materials science,
because $B_2^0$ can be evaluated with relatively high accuracy compared with the other higher-order CEF coefficients.

\begin{acknowledgments}
We would like to thank Prof. H. Kato, Dr. Y. Toga, and Mr. D. Suzuki
for useful discussions and information.
This work was supported by
JSPS KAKENHI Grant
Nos. 16K06702,
16H02390,
16H04322,
and 17K14800.
\end{acknowledgments}

\bibliographystyle{apsrev4-1}
\bibliography{library}

\begin{thebibliography}{63}%
\makeatletter
\providecommand \@ifxundefined [1]{%
 \@ifx{#1\undefined}
}%
\providecommand \@ifnum [1]{%
 \ifnum #1\expandafter \@firstoftwo
 \else \expandafter \@secondoftwo
 \fi
}%
\providecommand \@ifx [1]{%
 \ifx #1\expandafter \@firstoftwo
 \else \expandafter \@secondoftwo
 \fi
}%
\providecommand \natexlab [1]{#1}%
\providecommand \enquote  [1]{``#1''}%
\providecommand \bibnamefont  [1]{#1}%
\providecommand \bibfnamefont [1]{#1}%
\providecommand \citenamefont [1]{#1}%
\providecommand \href@noop [0]{\@secondoftwo}%
\providecommand \href [0]{\begingroup \@sanitize@url \@href}%
\providecommand \@href[1]{\@@startlink{#1}\@@href}%
\providecommand \@@href[1]{\endgroup#1\@@endlink}%
\providecommand \@sanitize@url [0]{\catcode `\\12\catcode `\$12\catcode
  `\&12\catcode `\#12\catcode `\^12\catcode `\_12\catcode `\%12\relax}%
\providecommand \@@startlink[1]{}%
\providecommand \@@endlink[0]{}%
\providecommand \url  [0]{\begingroup\@sanitize@url \@url }%
\providecommand \@url [1]{\endgroup\@href {#1}{\urlprefix }}%
\providecommand \urlprefix  [0]{URL }%
\providecommand \Eprint [0]{\href }%
\providecommand \doibase [0]{http://dx.doi.org/}%
\providecommand \selectlanguage [0]{\@gobble}%
\providecommand \bibinfo  [0]{\@secondoftwo}%
\providecommand \bibfield  [0]{\@secondoftwo}%
\providecommand \translation [1]{[#1]}%
\providecommand \BibitemOpen [0]{}%
\providecommand \bibitemStop [0]{}%
\providecommand \bibitemNoStop [0]{.\EOS\space}%
\providecommand \EOS [0]{\spacefactor3000\relax}%
\providecommand \BibitemShut  [1]{\csname bibitem#1\endcsname}%
\let\auto@bib@innerbib\@empty
%</preamble>
\bibitem [{\citenamefont {Herbst}(1991)}]{Herbst1991}%
  \BibitemOpen
  \bibfield  {author} {\bibinfo {author} {\bibfnamefont {J.~F.}\ \bibnamefont
  {Herbst}},\ }\href {\doibase 10.1103/RevModPhys.63.819} {\bibfield  {journal}
  {\bibinfo  {journal} {Rev. Mod. Phys.}\ }\textbf {\bibinfo {volume} {63}},\
  \bibinfo {pages} {819} (\bibinfo {year} {1991})}\BibitemShut {NoStop}%
\bibitem [{\citenamefont {Skomski}\ and\ \citenamefont
  {Coey}(1999)}]{Skomski1999}%
  \BibitemOpen
  \bibfield  {author} {\bibinfo {author} {\bibfnamefont {R.}~\bibnamefont
  {Skomski}}\ and\ \bibinfo {author} {\bibfnamefont {J.~M.~D.}\ \bibnamefont
  {Coey}},\ }\href@noop {} {\emph {\bibinfo {title} {{Permanent Magnetism}}}}\
  (\bibinfo  {publisher} {Institute of Physics Publishing},\ \bibinfo {address}
  {Bristol},\ \bibinfo {year} {1999})\BibitemShut {NoStop}%
\bibitem [{\citenamefont {Kuz'min}\ and\ \citenamefont
  {Tishin}(2007)}]{Kuzmin2007}%
  \BibitemOpen
  \bibfield  {author} {\bibinfo {author} {\bibfnamefont {M.~D.}\ \bibnamefont
  {Kuz'min}}\ and\ \bibinfo {author} {\bibfnamefont {A.~M.}\ \bibnamefont
  {Tishin}},\ }in\ \href {\doibase 10.1016/S1567-2719(07)17003-7} {\emph
  {\bibinfo {booktitle} {Handb. Magn. Mater.}}},\ \bibinfo {editor} {edited by\
  \bibinfo {editor} {\bibfnamefont {K.~H.~J.}\ \bibnamefont {Buschow}}}\
  (\bibinfo  {publisher} {Elsevier},\ \bibinfo {address} {Amsterdam},\ \bibinfo
  {year} {2007})\ p.\ \bibinfo {pages} {149}\BibitemShut {NoStop}%
\bibitem [{\citenamefont {Coey}(2010)}]{Coey2010}%
  \BibitemOpen
  \bibfield  {author} {\bibinfo {author} {\bibfnamefont {J.~M.~D.}\
  \bibnamefont {Coey}},\ }\href@noop {} {\emph {\bibinfo {title} {{Magnetism
  and Magnetic Materials}}}}\ (\bibinfo  {publisher} {Cambridge University
  Press},\ \bibinfo {address} {Cambridge},\ \bibinfo {year} {2010})\BibitemShut
  {NoStop}%
\bibitem [{\citenamefont {Hirosawa}\ \emph {et~al.}(2017)\citenamefont
  {Hirosawa}, \citenamefont {Nishino},\ and\ \citenamefont
  {Miyashita}}]{Hirosawa2017}%
  \BibitemOpen
  \bibfield  {author} {\bibinfo {author} {\bibfnamefont {S.}~\bibnamefont
  {Hirosawa}}, \bibinfo {author} {\bibfnamefont {M.}~\bibnamefont {Nishino}}, \
  and\ \bibinfo {author} {\bibfnamefont {S.}~\bibnamefont {Miyashita}},\ }\href
  {\doibase 10.1088/2043-6254/aa597c} {\bibfield  {journal} {\bibinfo
  {journal} {Adv. Nat. Sci. Nanosci. Nanotechnol.}\ }\textbf {\bibinfo {volume}
  {8}},\ \bibinfo {pages} {013002} (\bibinfo {year} {2017})}\BibitemShut
  {NoStop}%
\bibitem [{\citenamefont {Miyake}\ and\ \citenamefont
  {Akai}(2018)}]{Miyake2018}%
  \BibitemOpen
  \bibfield  {author} {\bibinfo {author} {\bibfnamefont {T.}~\bibnamefont
  {Miyake}}\ and\ \bibinfo {author} {\bibfnamefont {H.}~\bibnamefont {Akai}},\
  }\href {\doibase 10.7566/JPSJ.87.041009} {\bibfield  {journal} {\bibinfo
  {journal} {J. Phys. Soc. Jpn.}\ }\textbf {\bibinfo {volume} {87}},\ \bibinfo
  {pages} {041009} (\bibinfo {year} {2018})}\BibitemShut {NoStop}%
\bibitem [{\citenamefont {Asselin}\ \emph {et~al.}(2010)\citenamefont
  {Asselin}, \citenamefont {Evans}, \citenamefont {Barker}, \citenamefont
  {Chantrell}, \citenamefont {Yanes}, \citenamefont {Chubykalo-Fesenko},
  \citenamefont {Hinzke},\ and\ \citenamefont {Nowak}}]{Asselin2010}%
  \BibitemOpen
  \bibfield  {author} {\bibinfo {author} {\bibfnamefont {P.}~\bibnamefont
  {Asselin}}, \bibinfo {author} {\bibfnamefont {R.~F.~L.}\ \bibnamefont
  {Evans}}, \bibinfo {author} {\bibfnamefont {J.}~\bibnamefont {Barker}},
  \bibinfo {author} {\bibfnamefont {R.~W.}\ \bibnamefont {Chantrell}}, \bibinfo
  {author} {\bibfnamefont {R.}~\bibnamefont {Yanes}}, \bibinfo {author}
  {\bibfnamefont {O.}~\bibnamefont {Chubykalo-Fesenko}}, \bibinfo {author}
  {\bibfnamefont {D.}~\bibnamefont {Hinzke}}, \ and\ \bibinfo {author}
  {\bibfnamefont {U.}~\bibnamefont {Nowak}},\ }\href {\doibase
  10.1103/PhysRevB.82.054415} {\bibfield  {journal} {\bibinfo  {journal} {Phys.
  Rev. B}\ }\textbf {\bibinfo {volume} {82}},\ \bibinfo {pages} {054415}
  (\bibinfo {year} {2010})}\BibitemShut {NoStop}%
\bibitem [{\citenamefont {Nishino}\ and\ \citenamefont
  {Miyashita}(2015)}]{Nishino2015}%
  \BibitemOpen
  \bibfield  {author} {\bibinfo {author} {\bibfnamefont {M.}~\bibnamefont
  {Nishino}}\ and\ \bibinfo {author} {\bibfnamefont {S.}~\bibnamefont
  {Miyashita}},\ }\href {\doibase 10.1103/PhysRevB.91.134411} {\bibfield
  {journal} {\bibinfo  {journal} {Phys. Rev. B}\ }\textbf {\bibinfo {volume}
  {91}},\ \bibinfo {pages} {134411} (\bibinfo {year} {2015})}\BibitemShut
  {NoStop}%
\bibitem [{\citenamefont {Nishino}\ and\ \citenamefont
  {Miyashita}(2018)}]{Nishino2018}%
  \BibitemOpen
  \bibfield  {author} {\bibinfo {author} {\bibfnamefont {M.}~\bibnamefont
  {Nishino}}\ and\ \bibinfo {author} {\bibfnamefont {S.}~\bibnamefont
  {Miyashita}},\ }\href {\doibase 10.1103/PhysRevB.97.019904} {\bibfield
  {journal} {\bibinfo  {journal} {Phys. Rev. B}\ }\textbf {\bibinfo {volume}
  {97}},\ \bibinfo {pages} {019904(E)} (\bibinfo {year} {2018})}\BibitemShut
  {NoStop}%
\bibitem [{\citenamefont {Miura}\ \emph {et~al.}(2014)\citenamefont {Miura},
  \citenamefont {Tsuchiura},\ and\ \citenamefont {Yoshioka}}]{Miura2014d}%
  \BibitemOpen
  \bibfield  {author} {\bibinfo {author} {\bibfnamefont {Y.}~\bibnamefont
  {Miura}}, \bibinfo {author} {\bibfnamefont {H.}~\bibnamefont {Tsuchiura}}, \
  and\ \bibinfo {author} {\bibfnamefont {T.}~\bibnamefont {Yoshioka}},\ }\href
  {\doibase 10.1063/1.4869061} {\bibfield  {journal} {\bibinfo  {journal} {J.
  Appl. Phys.}\ }\textbf {\bibinfo {volume} {115}},\ \bibinfo {pages} {17A765}
  (\bibinfo {year} {2014})}\BibitemShut {NoStop}%
\bibitem [{\citenamefont {Yoshioka}\ \emph {et~al.}(2015)\citenamefont
  {Yoshioka}, \citenamefont {Tsuchiura},\ and\ \citenamefont
  {Nov{\'{a}}k}}]{Yoshioka2015}%
  \BibitemOpen
  \bibfield  {author} {\bibinfo {author} {\bibfnamefont {T.}~\bibnamefont
  {Yoshioka}}, \bibinfo {author} {\bibfnamefont {H.}~\bibnamefont {Tsuchiura}},
  \ and\ \bibinfo {author} {\bibfnamefont {P.}~\bibnamefont {Nov{\'{a}}k}},\
  }\href {\doibase 10.1179/1432891715Z.0000000001416} {\bibfield  {journal}
  {\bibinfo  {journal} {Mater. Res. Innov.}\ }\textbf {\bibinfo {volume}
  {19}},\ \bibinfo {pages} {S3} (\bibinfo {year} {2015})}\BibitemShut {NoStop}%
\bibitem [{\citenamefont {Tatetsu}\ \emph {et~al.}(2016)\citenamefont
  {Tatetsu}, \citenamefont {Tsuneyuki},\ and\ \citenamefont
  {Gohda}}]{Tatetsu2016}%
  \BibitemOpen
  \bibfield  {author} {\bibinfo {author} {\bibfnamefont {Y.}~\bibnamefont
  {Tatetsu}}, \bibinfo {author} {\bibfnamefont {S.}~\bibnamefont {Tsuneyuki}},
  \ and\ \bibinfo {author} {\bibfnamefont {Y.}~\bibnamefont {Gohda}},\ }\href
  {\doibase 10.1103/PhysRevApplied.6.064029} {\bibfield  {journal} {\bibinfo
  {journal} {Phys. Rev. Appl.}\ }\textbf {\bibinfo {volume} {6}},\ \bibinfo
  {pages} {064029} (\bibinfo {year} {2016})}\BibitemShut {NoStop}%
\bibitem [{\citenamefont {Tatetsu}\ \emph {et~al.}(2018)\citenamefont
  {Tatetsu}, \citenamefont {Harashima}, \citenamefont {Miyake},\ and\
  \citenamefont {Gohda}}]{Tatetsu2018a}%
  \BibitemOpen
  \bibfield  {author} {\bibinfo {author} {\bibfnamefont {Y.}~\bibnamefont
  {Tatetsu}}, \bibinfo {author} {\bibfnamefont {Y.}~\bibnamefont {Harashima}},
  \bibinfo {author} {\bibfnamefont {T.}~\bibnamefont {Miyake}}, \ and\ \bibinfo
  {author} {\bibfnamefont {Y.}~\bibnamefont {Gohda}},\ }\href {\doibase
  10.1103/PhysRevMaterials.2.074410} {\bibfield  {journal} {\bibinfo  {journal}
  {Phys. Rev. Mater.}\ }\textbf {\bibinfo {volume} {2}},\ \bibinfo {pages}
  {074410} (\bibinfo {year} {2018})}\BibitemShut {NoStop}%
\bibitem [{\citenamefont {Yoshioka}\ and\ \citenamefont
  {Tsuchiura}(2018)}]{Yoshioka2018}%
  \BibitemOpen
  \bibfield  {author} {\bibinfo {author} {\bibfnamefont {T.}~\bibnamefont
  {Yoshioka}}\ and\ \bibinfo {author} {\bibfnamefont {H.}~\bibnamefont
  {Tsuchiura}},\ }\href {\doibase 10.1063/1.5018875} {\bibfield  {journal}
  {\bibinfo  {journal} {Appl. Phys. Lett.}\ }\textbf {\bibinfo {volume}
  {112}},\ \bibinfo {pages} {162405} (\bibinfo {year} {2018})}\BibitemShut
  {NoStop}%
\bibitem [{\citenamefont {Tsuchiura}\ \emph {et~al.}(2018)\citenamefont
  {Tsuchiura}, \citenamefont {Yoshioka},\ and\ \citenamefont
  {Nov{\'{a}}k}}]{Tsuchiura2018a}%
  \BibitemOpen
  \bibfield  {author} {\bibinfo {author} {\bibfnamefont {H.}~\bibnamefont
  {Tsuchiura}}, \bibinfo {author} {\bibfnamefont {T.}~\bibnamefont {Yoshioka}},
  \ and\ \bibinfo {author} {\bibfnamefont {P.}~\bibnamefont {Nov{\'{a}}k}},\
  }\href {\doibase 10.1016/j.scriptamat.2018.04.023} {\bibfield  {journal}
  {\bibinfo  {journal} {Scr. Mater.}\ }\textbf {\bibinfo {volume} {154}},\
  \bibinfo {pages} {248} (\bibinfo {year} {2018})}\BibitemShut {NoStop}%
\bibitem [{\citenamefont {Toga}\ \emph {et~al.}(2016)\citenamefont {Toga},
  \citenamefont {Matsumoto}, \citenamefont {Miyashita}, \citenamefont {Akai},
  \citenamefont {Doi}, \citenamefont {Miyake},\ and\ \citenamefont
  {Sakuma}}]{Toga2016}%
  \BibitemOpen
  \bibfield  {author} {\bibinfo {author} {\bibfnamefont {Y.}~\bibnamefont
  {Toga}}, \bibinfo {author} {\bibfnamefont {M.}~\bibnamefont {Matsumoto}},
  \bibinfo {author} {\bibfnamefont {S.}~\bibnamefont {Miyashita}}, \bibinfo
  {author} {\bibfnamefont {H.}~\bibnamefont {Akai}}, \bibinfo {author}
  {\bibfnamefont {S.}~\bibnamefont {Doi}}, \bibinfo {author} {\bibfnamefont
  {T.}~\bibnamefont {Miyake}}, \ and\ \bibinfo {author} {\bibfnamefont
  {A.}~\bibnamefont {Sakuma}},\ }\href {\doibase 10.1103/PhysRevB.94.174433}
  {\bibfield  {journal} {\bibinfo  {journal} {Phys. Rev. B}\ }\textbf {\bibinfo
  {volume} {94}},\ \bibinfo {pages} {174433} (\bibinfo {year}
  {2016})}\BibitemShut {NoStop}%
\bibitem [{\citenamefont {Matsumoto}\ \emph {et~al.}(2016)\citenamefont
  {Matsumoto}, \citenamefont {Akai}, \citenamefont {Harashima}, \citenamefont
  {Doi},\ and\ \citenamefont {Miyake}}]{Matsumoto2016}%
  \BibitemOpen
  \bibfield  {author} {\bibinfo {author} {\bibfnamefont {M.}~\bibnamefont
  {Matsumoto}}, \bibinfo {author} {\bibfnamefont {H.}~\bibnamefont {Akai}},
  \bibinfo {author} {\bibfnamefont {Y.}~\bibnamefont {Harashima}}, \bibinfo
  {author} {\bibfnamefont {S.}~\bibnamefont {Doi}}, \ and\ \bibinfo {author}
  {\bibfnamefont {T.}~\bibnamefont {Miyake}},\ }\href {\doibase
  10.1063/1.4952989} {\bibfield  {journal} {\bibinfo  {journal} {J. Appl.
  Phys.}\ }\textbf {\bibinfo {volume} {119}},\ \bibinfo {pages} {213901}
  (\bibinfo {year} {2016})}\BibitemShut {NoStop}%
\bibitem [{\citenamefont {Nishino}\ \emph {et~al.}(2017)\citenamefont
  {Nishino}, \citenamefont {Toga}, \citenamefont {Miyashita}, \citenamefont
  {Akai}, \citenamefont {Sakuma},\ and\ \citenamefont
  {Hirosawa}}]{Nishino2017}%
  \BibitemOpen
  \bibfield  {author} {\bibinfo {author} {\bibfnamefont {M.}~\bibnamefont
  {Nishino}}, \bibinfo {author} {\bibfnamefont {Y.}~\bibnamefont {Toga}},
  \bibinfo {author} {\bibfnamefont {S.}~\bibnamefont {Miyashita}}, \bibinfo
  {author} {\bibfnamefont {H.}~\bibnamefont {Akai}}, \bibinfo {author}
  {\bibfnamefont {A.}~\bibnamefont {Sakuma}}, \ and\ \bibinfo {author}
  {\bibfnamefont {S.}~\bibnamefont {Hirosawa}},\ }\href {\doibase
  10.1103/PhysRevB.95.094429} {\bibfield  {journal} {\bibinfo  {journal} {Phys.
  Rev. B}\ }\textbf {\bibinfo {volume} {95}},\ \bibinfo {pages} {094429}
  (\bibinfo {year} {2017})}\BibitemShut {NoStop}%
\bibitem [{\citenamefont {Toga}\ \emph {et~al.}(2018)\citenamefont {Toga},
  \citenamefont {Nishino}, \citenamefont {Miyashita}, \citenamefont {Miyake},\
  and\ \citenamefont {Sakuma}}]{Toga2018}%
  \BibitemOpen
  \bibfield  {author} {\bibinfo {author} {\bibfnamefont {Y.}~\bibnamefont
  {Toga}}, \bibinfo {author} {\bibfnamefont {M.}~\bibnamefont {Nishino}},
  \bibinfo {author} {\bibfnamefont {S.}~\bibnamefont {Miyashita}}, \bibinfo
  {author} {\bibfnamefont {T.}~\bibnamefont {Miyake}}, \ and\ \bibinfo {author}
  {\bibfnamefont {A.}~\bibnamefont {Sakuma}},\ }\href {\doibase
  10.1103/PhysRevB.98.054418} {\bibfield  {journal} {\bibinfo  {journal} {Phys.
  Rev. B}\ }\textbf {\bibinfo {volume} {98}},\ \bibinfo {pages} {054418}
  (\bibinfo {year} {2018})}\BibitemShut {NoStop}%
\bibitem [{\citenamefont {Westmoreland}\ \emph {et~al.}(2018)\citenamefont
  {Westmoreland}, \citenamefont {Evans}, \citenamefont {Hrkac}, \citenamefont
  {Schrefl}, \citenamefont {Zimanyi}, \citenamefont {Winklhofer}, \citenamefont
  {Sakuma}, \citenamefont {Yano}, \citenamefont {Kato}, \citenamefont {Shoji},
  \citenamefont {Manabe}, \citenamefont {Ito},\ and\ \citenamefont
  {Chantrell}}]{Westmoreland2018}%
  \BibitemOpen
  \bibfield  {author} {\bibinfo {author} {\bibfnamefont {S.~C.}\ \bibnamefont
  {Westmoreland}}, \bibinfo {author} {\bibfnamefont {R.~F.}\ \bibnamefont
  {Evans}}, \bibinfo {author} {\bibfnamefont {G.}~\bibnamefont {Hrkac}},
  \bibinfo {author} {\bibfnamefont {T.}~\bibnamefont {Schrefl}}, \bibinfo
  {author} {\bibfnamefont {G.~T.}\ \bibnamefont {Zimanyi}}, \bibinfo {author}
  {\bibfnamefont {M.}~\bibnamefont {Winklhofer}}, \bibinfo {author}
  {\bibfnamefont {N.}~\bibnamefont {Sakuma}}, \bibinfo {author} {\bibfnamefont
  {M.}~\bibnamefont {Yano}}, \bibinfo {author} {\bibfnamefont {A.}~\bibnamefont
  {Kato}}, \bibinfo {author} {\bibfnamefont {T.}~\bibnamefont {Shoji}},
  \bibinfo {author} {\bibfnamefont {A.}~\bibnamefont {Manabe}}, \bibinfo
  {author} {\bibfnamefont {M.}~\bibnamefont {Ito}}, \ and\ \bibinfo {author}
  {\bibfnamefont {R.~W.}\ \bibnamefont {Chantrell}},\ }\href {\doibase
  10.1016/j.scriptamat.2018.04.019} {\bibfield  {journal} {\bibinfo  {journal}
  {Scr. Mater.}\ }\textbf {\bibinfo {volume} {148}},\ \bibinfo {pages} {56}
  (\bibinfo {year} {2018})}\BibitemShut {NoStop}%
\bibitem [{\citenamefont {Skomski}(1998)}]{Skomski1998}%
  \BibitemOpen
  \bibfield  {author} {\bibinfo {author} {\bibfnamefont {R.}~\bibnamefont
  {Skomski}},\ }\href@noop {} {\bibfield  {journal} {\bibinfo  {journal} {J.
  Appl. Phys.}\ }\textbf {\bibinfo {volume} {83}},\ \bibinfo {pages} {6724}
  (\bibinfo {year} {1998})}\BibitemShut {NoStop}%
\bibitem [{\citenamefont {Skomski}\ \emph {et~al.}(2006)\citenamefont
  {Skomski}, \citenamefont {Mryasov}, \citenamefont {Zhou},\ and\ \citenamefont
  {Sellmyer}}]{Skomski2006}%
  \BibitemOpen
  \bibfield  {author} {\bibinfo {author} {\bibfnamefont {R.}~\bibnamefont
  {Skomski}}, \bibinfo {author} {\bibfnamefont {O.~N.}\ \bibnamefont
  {Mryasov}}, \bibinfo {author} {\bibfnamefont {J.}~\bibnamefont {Zhou}}, \
  and\ \bibinfo {author} {\bibfnamefont {D.~J.}\ \bibnamefont {Sellmyer}},\
  }\href {\doibase 10.1063/1.2176892} {\bibfield  {journal} {\bibinfo
  {journal} {J. Appl. Phys.}\ }\textbf {\bibinfo {volume} {99}},\ \bibinfo
  {pages} {6} (\bibinfo {year} {2006})}\BibitemShut {NoStop}%
\bibitem [{\citenamefont {Skomski}\ and\ \citenamefont
  {Sellmyer}(2009)}]{Skomski2009}%
  \BibitemOpen
  \bibfield  {author} {\bibinfo {author} {\bibfnamefont {R.}~\bibnamefont
  {Skomski}}\ and\ \bibinfo {author} {\bibfnamefont {D.~J.}\ \bibnamefont
  {Sellmyer}},\ }\href {\doibase 10.1016/S1002-0721(08)60314-2} {\bibfield
  {journal} {\bibinfo  {journal} {J. Rare Earths}\ }\textbf {\bibinfo {volume}
  {27}},\ \bibinfo {pages} {675} (\bibinfo {year} {2009})}\BibitemShut
  {NoStop}%
\bibitem [{\citenamefont {Miura}\ and\ \citenamefont
  {Sakuma}(2018)}]{Miura2018}%
  \BibitemOpen
  \bibfield  {author} {\bibinfo {author} {\bibfnamefont {D.}~\bibnamefont
  {Miura}}\ and\ \bibinfo {author} {\bibfnamefont {A.}~\bibnamefont {Sakuma}},\
  }\href {\doibase 10.1063/1.5021969} {\bibfield  {journal} {\bibinfo
  {journal} {AIP Adv.}\ }\textbf {\bibinfo {volume} {8}},\ \bibinfo {pages}
  {075114} (\bibinfo {year} {2018})}\BibitemShut {NoStop}%
\bibitem [{\citenamefont {Sasaki}\ \emph {et~al.}(2015)\citenamefont {Sasaki},
  \citenamefont {Miura},\ and\ \citenamefont {Sakuma}}]{Sasaki2015}%
  \BibitemOpen
  \bibfield  {author} {\bibinfo {author} {\bibfnamefont {R.}~\bibnamefont
  {Sasaki}}, \bibinfo {author} {\bibfnamefont {D.}~\bibnamefont {Miura}}, \
  and\ \bibinfo {author} {\bibfnamefont {A.}~\bibnamefont {Sakuma}},\ }\href
  {\doibase 10.7567/APEX.8.043004} {\bibfield  {journal} {\bibinfo  {journal}
  {Appl. Phys. Express}\ }\textbf {\bibinfo {volume} {8}},\ \bibinfo {pages}
  {043004} (\bibinfo {year} {2015})},\ \Eprint
  {http://arxiv.org/abs/1501.01782} {arXiv:1501.01782} \BibitemShut {NoStop}%
\bibitem [{\citenamefont {Miura}\ \emph {et~al.}(2015)\citenamefont {Miura},
  \citenamefont {Sasaki},\ and\ \citenamefont {Sakuma}}]{Miura2015}%
  \BibitemOpen
  \bibfield  {author} {\bibinfo {author} {\bibfnamefont {D.}~\bibnamefont
  {Miura}}, \bibinfo {author} {\bibfnamefont {R.}~\bibnamefont {Sasaki}}, \
  and\ \bibinfo {author} {\bibfnamefont {A.}~\bibnamefont {Sakuma}},\ }\href
  {\doibase 10.7567/APEX.8.113003} {\bibfield  {journal} {\bibinfo  {journal}
  {Appl. Phys. Express}\ }\textbf {\bibinfo {volume} {8}},\ \bibinfo {pages}
  {113003} (\bibinfo {year} {2015})}\BibitemShut {NoStop}%
\bibitem [{\citenamefont {Ito}\ \emph {et~al.}(2016)\citenamefont {Ito},
  \citenamefont {Yano}, \citenamefont {Dempsey},\ and\ \citenamefont
  {Givord}}]{Ito2016}%
  \BibitemOpen
  \bibfield  {author} {\bibinfo {author} {\bibfnamefont {M.}~\bibnamefont
  {Ito}}, \bibinfo {author} {\bibfnamefont {M.}~\bibnamefont {Yano}}, \bibinfo
  {author} {\bibfnamefont {N.~M.}\ \bibnamefont {Dempsey}}, \ and\ \bibinfo
  {author} {\bibfnamefont {D.}~\bibnamefont {Givord}},\ }\href {\doibase
  10.1016/j.jmmm.2015.08.065} {\bibfield  {journal} {\bibinfo  {journal} {J.
  Magn. Magn. Mater.}\ }\textbf {\bibinfo {volume} {400}},\ \bibinfo {pages}
  {379} (\bibinfo {year} {2016})}\BibitemShut {NoStop}%
\bibitem [{\citenamefont {Skomski}\ \emph {et~al.}(2013)\citenamefont
  {Skomski}, \citenamefont {Kumar}, \citenamefont {Hadjipanayis},\ and\
  \citenamefont {Sellmyer}}]{Skomski2013}%
  \BibitemOpen
  \bibfield  {author} {\bibinfo {author} {\bibfnamefont {R.}~\bibnamefont
  {Skomski}}, \bibinfo {author} {\bibfnamefont {P.}~\bibnamefont {Kumar}},
  \bibinfo {author} {\bibfnamefont {G.~C.}\ \bibnamefont {Hadjipanayis}}, \
  and\ \bibinfo {author} {\bibfnamefont {D.~J.}\ \bibnamefont {Sellmyer}},\
  }\href@noop {} {\bibfield  {journal} {\bibinfo  {journal} {IEEE Trans.
  Magn.}\ }\textbf {\bibinfo {volume} {49}},\ \bibinfo {pages} {3229} (\bibinfo
  {year} {2013})}\BibitemShut {NoStop}%
\bibitem [{\citenamefont {Hirosawa}\ \emph {et~al.}(1986)\citenamefont
  {Hirosawa}, \citenamefont {Matsuura}, \citenamefont {Yamamoto}, \citenamefont
  {Fujimura}, \citenamefont {Sagawa},\ and\ \citenamefont
  {Yamauchi}}]{Hirosawa1986}%
  \BibitemOpen
  \bibfield  {author} {\bibinfo {author} {\bibfnamefont {S.}~\bibnamefont
  {Hirosawa}}, \bibinfo {author} {\bibfnamefont {Y.}~\bibnamefont {Matsuura}},
  \bibinfo {author} {\bibfnamefont {H.}~\bibnamefont {Yamamoto}}, \bibinfo
  {author} {\bibfnamefont {S.}~\bibnamefont {Fujimura}}, \bibinfo {author}
  {\bibfnamefont {M.}~\bibnamefont {Sagawa}}, \ and\ \bibinfo {author}
  {\bibfnamefont {H.}~\bibnamefont {Yamauchi}},\ }\href {\doibase
  10.1063/1.336611} {\bibfield  {journal} {\bibinfo  {journal} {J. Appl.
  Phys.}\ }\textbf {\bibinfo {volume} {59}},\ \bibinfo {pages} {873} (\bibinfo
  {year} {1986})}\BibitemShut {NoStop}%
\bibitem [{\citenamefont {Durst}\ and\ \citenamefont
  {Kronmuller}(1986)}]{Durst1986}%
  \BibitemOpen
  \bibfield  {author} {\bibinfo {author} {\bibfnamefont {K.~D.}\ \bibnamefont
  {Durst}}\ and\ \bibinfo {author} {\bibfnamefont {H.}~\bibnamefont
  {Kronmuller}},\ }\href@noop {} {\bibfield  {journal} {\bibinfo  {journal} {J.
  Magn. Magn. Mater.}\ }\textbf {\bibinfo {volume} {59}},\ \bibinfo {pages}
  {86} (\bibinfo {year} {1986})}\BibitemShut {NoStop}%
\bibitem [{\citenamefont {Sagawa}\ \emph {et~al.}(1987)\citenamefont {Sagawa},
  \citenamefont {Hirosawa}, \citenamefont {Yamamoto}, \citenamefont
  {Fujimura},\ and\ \citenamefont {Matsuura}}]{Sagawa1987}%
  \BibitemOpen
  \bibfield  {author} {\bibinfo {author} {\bibfnamefont {M.}~\bibnamefont
  {Sagawa}}, \bibinfo {author} {\bibfnamefont {S.}~\bibnamefont {Hirosawa}},
  \bibinfo {author} {\bibfnamefont {H.}~\bibnamefont {Yamamoto}}, \bibinfo
  {author} {\bibfnamefont {S.}~\bibnamefont {Fujimura}}, \ and\ \bibinfo
  {author} {\bibfnamefont {Y.}~\bibnamefont {Matsuura}},\ }\href@noop {}
  {\bibfield  {journal} {\bibinfo  {journal} {Jpn. J. Appl. Phys.}\ }\textbf
  {\bibinfo {volume} {26}},\ \bibinfo {pages} {785} (\bibinfo {year}
  {1987})}\BibitemShut {NoStop}%
\bibitem [{\citenamefont {Cadogan}\ \emph {et~al.}(1988)\citenamefont
  {Cadogan}, \citenamefont {Gavigan}, \citenamefont {Givord},\ and\
  \citenamefont {Li}}]{Cadogan1988}%
  \BibitemOpen
  \bibfield  {author} {\bibinfo {author} {\bibfnamefont {J.~M.}\ \bibnamefont
  {Cadogan}}, \bibinfo {author} {\bibfnamefont {J.~P.}\ \bibnamefont
  {Gavigan}}, \bibinfo {author} {\bibfnamefont {D.}~\bibnamefont {Givord}}, \
  and\ \bibinfo {author} {\bibfnamefont {H.~S.}\ \bibnamefont {Li}},\
  }\href@noop {} {\bibfield  {journal} {\bibinfo  {journal} {J. Phys. F Met.
  Phys.}\ }\textbf {\bibinfo {volume} {18}},\ \bibinfo {pages} {779} (\bibinfo
  {year} {1988})}\BibitemShut {NoStop}%
\bibitem [{\citenamefont {Kuz'min}(1995)}]{Kuzmin1995}%
  \BibitemOpen
  \bibfield  {author} {\bibinfo {author} {\bibfnamefont {M.~D.}\ \bibnamefont
  {Kuz'min}},\ }\href@noop {} {\bibfield  {journal} {\bibinfo  {journal} {Phys.
  Rev. B}\ }\textbf {\bibinfo {volume} {51}},\ \bibinfo {pages} {8904}
  (\bibinfo {year} {1995})}\BibitemShut {NoStop}%
\bibitem [{\citenamefont {Hirosawa}\ and\ \citenamefont
  {Sagawa}(1985)}]{Hirosawa1985a}%
  \BibitemOpen
  \bibfield  {author} {\bibinfo {author} {\bibfnamefont {S.}~\bibnamefont
  {Hirosawa}}\ and\ \bibinfo {author} {\bibfnamefont {M.}~\bibnamefont
  {Sagawa}},\ }\href {\doibase 10.1016/0038-1098(85)90009-2} {\bibfield
  {journal} {\bibinfo  {journal} {Solid State Commun.}\ }\textbf {\bibinfo
  {volume} {54}},\ \bibinfo {pages} {335} (\bibinfo {year} {1985})}\BibitemShut
  {NoStop}%
\bibitem [{\citenamefont {Hirosawa}\ \emph {et~al.}(1985)\citenamefont
  {Hirosawa}, \citenamefont {Matsuura}, \citenamefont {Yamamoto}, \citenamefont
  {Fujimura}, \citenamefont {Sagawa},\ and\ \citenamefont
  {Yamauchi}}]{Hirosawa1985}%
  \BibitemOpen
  \bibfield  {author} {\bibinfo {author} {\bibfnamefont {S.}~\bibnamefont
  {Hirosawa}}, \bibinfo {author} {\bibfnamefont {Y.}~\bibnamefont {Matsuura}},
  \bibinfo {author} {\bibfnamefont {H.}~\bibnamefont {Yamamoto}}, \bibinfo
  {author} {\bibfnamefont {S.}~\bibnamefont {Fujimura}}, \bibinfo {author}
  {\bibfnamefont {M.}~\bibnamefont {Sagawa}}, \ and\ \bibinfo {author}
  {\bibfnamefont {H.}~\bibnamefont {Yamauchi}},\ }\href {\doibase
  10.1143/JJAP.24.L803} {\bibfield  {journal} {\bibinfo  {journal} {Jpn. J.
  Appl. Phys.}\ }\textbf {\bibinfo {volume} {24}},\ \bibinfo {pages} {L803}
  (\bibinfo {year} {1985})}\BibitemShut {NoStop}%
\bibitem [{\citenamefont {Yamada}\ \emph {et~al.}(1986)\citenamefont {Yamada},
  \citenamefont {Tokuhara}, \citenamefont {Ono}, \citenamefont {Sagawa},\ and\
  \citenamefont {Matsuura}}]{Yamada1986}%
  \BibitemOpen
  \bibfield  {author} {\bibinfo {author} {\bibfnamefont {O.}~\bibnamefont
  {Yamada}}, \bibinfo {author} {\bibfnamefont {H.}~\bibnamefont {Tokuhara}},
  \bibinfo {author} {\bibfnamefont {F.}~\bibnamefont {Ono}}, \bibinfo {author}
  {\bibfnamefont {M.}~\bibnamefont {Sagawa}}, \ and\ \bibinfo {author}
  {\bibfnamefont {Y.}~\bibnamefont {Matsuura}},\ }\href {\doibase
  10.1016/0304-8853(86)90718-3} {\bibfield  {journal} {\bibinfo  {journal} {J.
  Magn. Magn. Mater.}\ }\textbf {\bibinfo {volume} {54-57}},\ \bibinfo {pages}
  {585} (\bibinfo {year} {1986})}\BibitemShut {NoStop}%
\bibitem [{\citenamefont {Grossinger}\ \emph {et~al.}(1986)\citenamefont
  {Grossinger}, \citenamefont {Sun}, \citenamefont {Eibler}, \citenamefont
  {Buschow},\ and\ \citenamefont {Kirchmayr}}]{Grossinger1986}%
  \BibitemOpen
  \bibfield  {author} {\bibinfo {author} {\bibfnamefont {R.}~\bibnamefont
  {Grossinger}}, \bibinfo {author} {\bibfnamefont {X.~K.}\ \bibnamefont {Sun}},
  \bibinfo {author} {\bibfnamefont {R.}~\bibnamefont {Eibler}}, \bibinfo
  {author} {\bibfnamefont {K.~H.~J.}\ \bibnamefont {Buschow}}, \ and\ \bibinfo
  {author} {\bibfnamefont {H.~R.}\ \bibnamefont {Kirchmayr}},\ }\href {\doibase
  10.1016/0304-8853(86)90122-8} {\bibfield  {journal} {\bibinfo  {journal} {J.
  Magn. Magn. Mater.}\ }\textbf {\bibinfo {volume} {58}},\ \bibinfo {pages}
  {55} (\bibinfo {year} {1986})}\BibitemShut {NoStop}%
\bibitem [{\citenamefont {Hirosawa}\ \emph {et~al.}(1987)\citenamefont
  {Hirosawa}, \citenamefont {Matsuura}, \citenamefont {Yamamoto}, \citenamefont
  {Fujimura}, \citenamefont {Sagawa},\ and\ \citenamefont
  {Yamauchi}}]{Hirosawa1987}%
  \BibitemOpen
  \bibfield  {author} {\bibinfo {author} {\bibfnamefont {S.}~\bibnamefont
  {Hirosawa}}, \bibinfo {author} {\bibfnamefont {Y.}~\bibnamefont {Matsuura}},
  \bibinfo {author} {\bibfnamefont {H.}~\bibnamefont {Yamamoto}}, \bibinfo
  {author} {\bibfnamefont {S.}~\bibnamefont {Fujimura}}, \bibinfo {author}
  {\bibfnamefont {M.}~\bibnamefont {Sagawa}}, \ and\ \bibinfo {author}
  {\bibfnamefont {H.}~\bibnamefont {Yamauchi}},\ }\href {\doibase
  10.1063/1.336611} {\bibfield  {journal} {\bibinfo  {journal} {J. Appl.
  Phys.}\ }\textbf {\bibinfo {volume} {61}},\ \bibinfo {pages} {3571} (\bibinfo
  {year} {1987})}\BibitemShut {NoStop}%
\bibitem [{\citenamefont {Otani}\ \emph {et~al.}(1987)\citenamefont {Otani},
  \citenamefont {Miyajima},\ and\ \citenamefont {Chikazumi}}]{Otani1987}%
  \BibitemOpen
  \bibfield  {author} {\bibinfo {author} {\bibfnamefont {Y.}~\bibnamefont
  {Otani}}, \bibinfo {author} {\bibfnamefont {H.}~\bibnamefont {Miyajima}}, \
  and\ \bibinfo {author} {\bibfnamefont {S.}~\bibnamefont {Chikazumi}},\ }\href
  {\doibase 10.1063/1.338745} {\bibfield  {journal} {\bibinfo  {journal} {J.
  Appl. Phys.}\ }\textbf {\bibinfo {volume} {61}},\ \bibinfo {pages} {3436}
  (\bibinfo {year} {1987})}\BibitemShut {NoStop}%
\bibitem [{\citenamefont {Radwanski}\ and\ \citenamefont
  {Franse}(1989)}]{Radwanski1989}%
  \BibitemOpen
  \bibfield  {author} {\bibinfo {author} {\bibfnamefont {R.~J.}\ \bibnamefont
  {Radwanski}}\ and\ \bibinfo {author} {\bibfnamefont {J.~J.~M.}\ \bibnamefont
  {Franse}},\ }\href@noop {} {\bibfield  {journal} {\bibinfo  {journal} {J.
  Magn. Magn. Mater.}\ }\textbf {\bibinfo {volume} {80}},\ \bibinfo {pages}
  {14} (\bibinfo {year} {1989})}\BibitemShut {NoStop}%
\bibitem [{\citenamefont {Zener}(1954)}]{Zener1954}%
  \BibitemOpen
  \bibfield  {author} {\bibinfo {author} {\bibfnamefont {C.}~\bibnamefont
  {Zener}},\ }\href {http://link.aps.org/doi/10.1103/PhysRev.96.1335}
  {\bibfield  {journal} {\bibinfo  {journal} {Phys. Rev.}\ }\textbf {\bibinfo
  {volume} {96}},\ \bibinfo {pages} {1335} (\bibinfo {year}
  {1954})}\BibitemShut {NoStop}%
\bibitem [{\citenamefont {Akulov}(1936)}]{Akulov1936}%
  \BibitemOpen
  \bibfield  {author} {\bibinfo {author} {\bibfnamefont {N.}~\bibnamefont
  {Akulov}},\ }\href@noop {} {\bibfield  {journal} {\bibinfo  {journal} {Z.
  Phys.}\ }\textbf {\bibinfo {volume} {100}},\ \bibinfo {pages} {197} (\bibinfo
  {year} {1936})}\BibitemShut {NoStop}%
\bibitem [{\citenamefont {Callen}\ and\ \citenamefont
  {Callen}(1966)}]{Callen1966}%
  \BibitemOpen
  \bibfield  {author} {\bibinfo {author} {\bibfnamefont {H.~B.}\ \bibnamefont
  {Callen}}\ and\ \bibinfo {author} {\bibfnamefont {E.}~\bibnamefont
  {Callen}},\ }\href {\doibase 10.1016/0022-3697(66)90012-6} {\bibfield
  {journal} {\bibinfo  {journal} {J. Phys. Chem. Solids}\ }\textbf {\bibinfo
  {volume} {27}},\ \bibinfo {pages} {1271} (\bibinfo {year}
  {1966})}\BibitemShut {NoStop}%
\bibitem [{\citenamefont {Kuz'min}(1992)}]{Kuzmin1992}%
  \BibitemOpen
  \bibfield  {author} {\bibinfo {author} {\bibfnamefont {M.~D.}\ \bibnamefont
  {Kuz'min}},\ }\href {\doibase 10.1103/PhysRevB.46.8219} {\bibfield  {journal}
  {\bibinfo  {journal} {Phys. Rev. B}\ }\textbf {\bibinfo {volume} {46}},\
  \bibinfo {pages} {8219} (\bibinfo {year} {1992})}\BibitemShut {NoStop}%
\bibitem [{\citenamefont {Yamada}\ \emph {et~al.}(1988)\citenamefont {Yamada},
  \citenamefont {Kato}, \citenamefont {Yamamoto},\ and\ \citenamefont
  {Nakagawa}}]{Yamada1988}%
  \BibitemOpen
  \bibfield  {author} {\bibinfo {author} {\bibfnamefont {M.}~\bibnamefont
  {Yamada}}, \bibinfo {author} {\bibfnamefont {H.}~\bibnamefont {Kato}},
  \bibinfo {author} {\bibfnamefont {H.}~\bibnamefont {Yamamoto}}, \ and\
  \bibinfo {author} {\bibfnamefont {Y.}~\bibnamefont {Nakagawa}},\ }\href
  {\doibase 10.1103/PhysRevB.38.620} {\bibfield  {journal} {\bibinfo  {journal}
  {Phys. Rev. B}\ }\textbf {\bibinfo {volume} {38}},\ \bibinfo {pages} {620}
  (\bibinfo {year} {1988})}\BibitemShut {NoStop}%
\bibitem [{\citenamefont {Kuz'min}\ \emph {et~al.}(1995)\citenamefont
  {Kuz'min}, \citenamefont {Garcia}, \citenamefont {Plaza},\ and\ \citenamefont
  {Fruchart}}]{Kuz1995}%
  \BibitemOpen
  \bibfield  {author} {\bibinfo {author} {\bibfnamefont {M.~D.}\ \bibnamefont
  {Kuz'min}}, \bibinfo {author} {\bibfnamefont {L.~M.}\ \bibnamefont {Garcia}},
  \bibinfo {author} {\bibfnamefont {I.}~\bibnamefont {Plaza}}, \ and\ \bibinfo
  {author} {\bibfnamefont {D.}~\bibnamefont {Fruchart}},\ }\href@noop {}
  {\bibfield  {journal} {\bibinfo  {journal} {J. Magn. Magn. Mater.}\ }\textbf
  {\bibinfo {volume} {146}},\ \bibinfo {pages} {77} (\bibinfo {year}
  {1995})}\BibitemShut {NoStop}%
\bibitem [{\citenamefont {Kuz'min}(1996)}]{Kuzmin1996}%
  \BibitemOpen
  \bibfield  {author} {\bibinfo {author} {\bibfnamefont {M.~D.}\ \bibnamefont
  {Kuz'min}},\ }\href {\doibase 10.1016/0304-8853(95)00619-2} {\bibfield
  {journal} {\bibinfo  {journal} {J. Magn. Magn. Mater.}\ }\textbf {\bibinfo
  {volume} {154}},\ \bibinfo {pages} {333} (\bibinfo {year}
  {1996})}\BibitemShut {NoStop}%
\bibitem [{\citenamefont {Keffer}(1955)}]{Keffer1955}%
  \BibitemOpen
  \bibfield  {author} {\bibinfo {author} {\bibfnamefont {F.}~\bibnamefont
  {Keffer}},\ }\href@noop {} {\bibfield  {journal} {\bibinfo  {journal} {Phys.
  Rev.}\ }\textbf {\bibinfo {volume} {100}},\ \bibinfo {pages} {1692} (\bibinfo
  {year} {1955})}\BibitemShut {NoStop}%
\bibitem [{\citenamefont {Kazakov}\ and\ \citenamefont
  {Andreeva}(1970)}]{Kazakov1970}%
  \BibitemOpen
  \bibfield  {author} {\bibinfo {author} {\bibfnamefont {A.~A.}\ \bibnamefont
  {Kazakov}}\ and\ \bibinfo {author} {\bibfnamefont {R.~I.}\ \bibnamefont
  {Andreeva}},\ }\href@noop {} {\bibfield  {journal} {\bibinfo  {journal} {Sov.
  Phys. Solid State}\ }\textbf {\bibinfo {volume} {12}},\ \bibinfo {pages}
  {192} (\bibinfo {year} {1970})}\BibitemShut {NoStop}%
\bibitem [{\citenamefont {Haskel}\ \emph {et~al.}(2005)\citenamefont {Haskel},
  \citenamefont {Lang}, \citenamefont {Islam}, \citenamefont {Cady},
  \citenamefont {Srajer}, \citenamefont {van Veenendaal},\ and\ \citenamefont
  {Canfield}}]{Haskel2005}%
  \BibitemOpen
  \bibfield  {author} {\bibinfo {author} {\bibfnamefont {D.}~\bibnamefont
  {Haskel}}, \bibinfo {author} {\bibfnamefont {J.~C.}\ \bibnamefont {Lang}},
  \bibinfo {author} {\bibfnamefont {Z.}~\bibnamefont {Islam}}, \bibinfo
  {author} {\bibfnamefont {A.}~\bibnamefont {Cady}}, \bibinfo {author}
  {\bibfnamefont {G.}~\bibnamefont {Srajer}}, \bibinfo {author} {\bibfnamefont
  {M.}~\bibnamefont {van Veenendaal}}, \ and\ \bibinfo {author} {\bibfnamefont
  {P.~C.}\ \bibnamefont {Canfield}},\ }\href {\doibase
  10.1103/PhysRevLett.95.217207} {\bibfield  {journal} {\bibinfo  {journal}
  {Phys. Rev. Lett.}\ }\textbf {\bibinfo {volume} {95}},\ \bibinfo {pages}
  {217207} (\bibinfo {year} {2005})}\BibitemShut {NoStop}%
\bibitem [{\citenamefont {Kuz'min}(2005)}]{Kuzmin2005}%
  \BibitemOpen
  \bibfield  {author} {\bibinfo {author} {\bibfnamefont {M.~D.}\ \bibnamefont
  {Kuz'min}},\ }\href {\doibase 10.1103/PhysRevLett.94.107204} {\bibfield
  {journal} {\bibinfo  {journal} {Phys. Rev. Lett.}\ }\textbf {\bibinfo
  {volume} {94}},\ \bibinfo {pages} {107204} (\bibinfo {year}
  {2005})}\BibitemShut {NoStop}%
\bibitem [{\citenamefont {Kuz'min}\ \emph {et~al.}(2010)\citenamefont
  {Kuz'min}, \citenamefont {Givord},\ and\ \citenamefont
  {Skumryev}}]{Kuzmin2010}%
  \BibitemOpen
  \bibfield  {author} {\bibinfo {author} {\bibfnamefont {M.~D.}\ \bibnamefont
  {Kuz'min}}, \bibinfo {author} {\bibfnamefont {D.}~\bibnamefont {Givord}}, \
  and\ \bibinfo {author} {\bibfnamefont {V.}~\bibnamefont {Skumryev}},\ }\href
  {\doibase 10.1063/1.3443576} {\bibfield  {journal} {\bibinfo  {journal} {J.
  Appl. Phys.}\ }\textbf {\bibinfo {volume} {107}},\ \bibinfo {pages} {113924}
  (\bibinfo {year} {2010})}\BibitemShut {NoStop}%
\bibitem [{\citenamefont {{G{\'{o}}mez Eslava}}\ \emph
  {et~al.}(2016)\citenamefont {{G{\'{o}}mez Eslava}}, \citenamefont {Ito},
  \citenamefont {Yano}, \citenamefont {Dempsey},\ and\ \citenamefont
  {Givord}}]{GomezEslava2016}%
  \BibitemOpen
  \bibfield  {author} {\bibinfo {author} {\bibfnamefont {G.}~\bibnamefont
  {{G{\'{o}}mez Eslava}}}, \bibinfo {author} {\bibfnamefont {M.}~\bibnamefont
  {Ito}}, \bibinfo {author} {\bibfnamefont {M.}~\bibnamefont {Yano}}, \bibinfo
  {author} {\bibfnamefont {N.~M.}\ \bibnamefont {Dempsey}}, \ and\ \bibinfo
  {author} {\bibfnamefont {D.}~\bibnamefont {Givord}},\ }\href {\doibase
  10.1016/j.jsamd.2016.06.014} {\bibfield  {journal} {\bibinfo  {journal} {J.
  Sci. Adv. Mater. Devices}\ }\textbf {\bibinfo {volume} {1}},\ \bibinfo
  {pages} {158} (\bibinfo {year} {2016})}\BibitemShut {NoStop}%
\bibitem [{\citenamefont {Diop}\ \emph {et~al.}(2016)\citenamefont {Diop},
  \citenamefont {Kuz'min}, \citenamefont {Skokov}, \citenamefont {Karpenkov},\
  and\ \citenamefont {Gutfleisch}}]{Diop2016}%
  \BibitemOpen
  \bibfield  {author} {\bibinfo {author} {\bibfnamefont {L.~V.}\ \bibnamefont
  {Diop}}, \bibinfo {author} {\bibfnamefont {M.~D.}\ \bibnamefont {Kuz'min}},
  \bibinfo {author} {\bibfnamefont {K.~P.}\ \bibnamefont {Skokov}}, \bibinfo
  {author} {\bibfnamefont {D.~Y.}\ \bibnamefont {Karpenkov}}, \ and\ \bibinfo
  {author} {\bibfnamefont {O.}~\bibnamefont {Gutfleisch}},\ }\href {\doibase
  10.1103/PhysRevB.94.144413} {\bibfield  {journal} {\bibinfo  {journal} {Phys.
  Rev. B}\ }\textbf {\bibinfo {volume} {94}},\ \bibinfo {pages} {144413}
  (\bibinfo {year} {2016})}\BibitemShut {NoStop}%
\bibitem [{\citenamefont {Ozaki}\ \emph {et~al.}(2017)\citenamefont {Ozaki},
  \citenamefont {Miura},\ and\ \citenamefont {Sakuma}}]{Ozaki2017}%
  \BibitemOpen
  \bibfield  {author} {\bibinfo {author} {\bibfnamefont {D.}~\bibnamefont
  {Ozaki}}, \bibinfo {author} {\bibfnamefont {D.}~\bibnamefont {Miura}}, \ and\
  \bibinfo {author} {\bibfnamefont {A.}~\bibnamefont {Sakuma}},\ }\href@noop {}
  {\bibfield  {journal} {\bibinfo  {journal} {IEEE Trans. Magn.}\ }\textbf
  {\bibinfo {volume} {53}},\ \bibinfo {pages} {1300604} (\bibinfo {year}
  {2017})}\BibitemShut {NoStop}%
\bibitem [{Note1()}]{Note1}%
  \BibitemOpen
  \bibinfo {note} {$T_\protect \mathrm {C}=586$ K for the Nd$_2$Fe$_{14}$B
  magnet\cite {Hirosawa1986,Sagawa1987}, and the parameters included in
  $\protect \mathcal {H}^i(\theta ,\phi ;T)$ are listed in Ref. \protect
  \citenum {Yamada1988}.}\BibitemShut {Stop}%
\bibitem [{Note2()}]{Note2}%
  \BibitemOpen
  \bibinfo {note} {$T_\protect \mathrm {C}=598$ K for the Dy$_2$Fe$_{14}$B
  magnet\cite {Hirosawa1986,Sagawa1987}, and the parameters included in
  $\protect \mathcal {H}^i(\theta ,\phi ;T)$ are listed in Ref. \protect
  \citenum {Yamada1988}.}\BibitemShut {Stop}%
\bibitem [{\citenamefont {Kuz'min}\ and\ \citenamefont
  {Coey}(1994)}]{Kuzmin1994}%
  \BibitemOpen
  \bibfield  {author} {\bibinfo {author} {\bibfnamefont {M.~D.}\ \bibnamefont
  {Kuz'min}}\ and\ \bibinfo {author} {\bibfnamefont {J.~M.~D.}\ \bibnamefont
  {Coey}},\ }\href {\doibase 10.1103/PhysRevB.50.12533} {\bibfield  {journal}
  {\bibinfo  {journal} {Phys. Rev. B}\ }\textbf {\bibinfo {volume} {50}},\
  \bibinfo {pages} {12533} (\bibinfo {year} {1994})}\BibitemShut {NoStop}%
\bibitem [{\citenamefont {Magnani}\ \emph {et~al.}(2000)\citenamefont
  {Magnani}, \citenamefont {Amoretti}, \citenamefont {Paoluzi},\ and\
  \citenamefont {Pareti}}]{Magnani2000}%
  \BibitemOpen
  \bibfield  {author} {\bibinfo {author} {\bibfnamefont {N.}~\bibnamefont
  {Magnani}}, \bibinfo {author} {\bibfnamefont {G.}~\bibnamefont {Amoretti}},
  \bibinfo {author} {\bibfnamefont {A.}~\bibnamefont {Paoluzi}}, \ and\
  \bibinfo {author} {\bibfnamefont {L.}~\bibnamefont {Pareti}},\ }\href
  {\doibase 10.1103/PhysRevB.62.9453} {\bibfield  {journal} {\bibinfo
  {journal} {Phys. Rev. B}\ }\textbf {\bibinfo {volume} {62}},\ \bibinfo
  {pages} {9453} (\bibinfo {year} {2000})}\BibitemShut {NoStop}%
\bibitem [{\citenamefont {Magnani}\ \emph {et~al.}(2001)\citenamefont
  {Magnani}, \citenamefont {Amoretti}, \citenamefont {Paoluzi},\ and\
  \citenamefont {Pareti}}]{Magnani2001}%
  \BibitemOpen
  \bibfield  {author} {\bibinfo {author} {\bibfnamefont {N.}~\bibnamefont
  {Magnani}}, \bibinfo {author} {\bibfnamefont {G.}~\bibnamefont {Amoretti}},
  \bibinfo {author} {\bibfnamefont {A.}~\bibnamefont {Paoluzi}}, \ and\
  \bibinfo {author} {\bibfnamefont {L.}~\bibnamefont {Pareti}},\ }\href
  {\doibase 10.1109/20.951228} {\bibfield  {journal} {\bibinfo  {journal} {IEEE
  Trans. Magn.}\ }\textbf {\bibinfo {volume} {37}},\ \bibinfo {pages} {2540}
  (\bibinfo {year} {2001})}\BibitemShut {NoStop}%
\bibitem [{\citenamefont {Kuz'min}(2002)}]{Kuzmin2002}%
  \BibitemOpen
  \bibfield  {author} {\bibinfo {author} {\bibfnamefont {M.~D.}\ \bibnamefont
  {Kuz'min}},\ }\href {\doibase 10.1063/1.1513876} {\bibfield  {journal}
  {\bibinfo  {journal} {J. Appl. Phys.}\ }\textbf {\bibinfo {volume} {92}},\
  \bibinfo {pages} {6693} (\bibinfo {year} {2002})}\BibitemShut {NoStop}%
\bibitem [{\citenamefont {Magnani}\ \emph {et~al.}(2003)\citenamefont
  {Magnani}, \citenamefont {Carretta}, \citenamefont {Liviotti},\ and\
  \citenamefont {Amoretti}}]{Magnani2003}%
  \BibitemOpen
  \bibfield  {author} {\bibinfo {author} {\bibfnamefont {N.}~\bibnamefont
  {Magnani}}, \bibinfo {author} {\bibfnamefont {S.}~\bibnamefont {Carretta}},
  \bibinfo {author} {\bibfnamefont {E.}~\bibnamefont {Liviotti}}, \ and\
  \bibinfo {author} {\bibfnamefont {G.}~\bibnamefont {Amoretti}},\ }\href
  {\doibase 10.1103/PhysRevB.67.144411} {\bibfield  {journal} {\bibinfo
  {journal} {Phys. Rev. B}\ }\textbf {\bibinfo {volume} {67}},\ \bibinfo
  {pages} {144411} (\bibinfo {year} {2003})}\BibitemShut {NoStop}%
\bibitem [{\citenamefont {Stevens}(1952)}]{Stevens1952}%
  \BibitemOpen
  \bibfield  {author} {\bibinfo {author} {\bibfnamefont {K.~W.~H.}\
  \bibnamefont {Stevens}},\ }\href {\doibase 10.1088/0370-1298/65/3/308}
  {\bibfield  {journal} {\bibinfo  {journal} {Proc. Phys. Soc. A}\ }\textbf
  {\bibinfo {volume} {65}},\ \bibinfo {pages} {209} (\bibinfo {year}
  {1952})}\BibitemShut {NoStop}%
\end{thebibliography}%

\end{document}